\pdfoutput=1
\documentclass[sigconf,
    authorversion,
    nonacm
]{acmart}

\usepackage[english]{babel}
\usepackage[utf8]{inputenc} %
\usepackage{blindtext}
\usepackage{comment}

\usepackage{url}
\usepackage{multirow} %
\usepackage{xspace}
\usepackage{color, colortbl} 
\usepackage{cleveref} %
\usepackage{graphicx}
\usepackage{pifont} %
\usepackage{tikz}
\usepackage{bm} %
\usepackage{todonotes}
\usepackage{caption}
\usepackage{soul} %
\usepackage{ifthen}
\usepackage{algorithm}
\usepackage[noend]{algpseudocode} %
\usepackage{subcaption} %
\usepackage[nolist]{acronym}
\usepackage{listings}
\usepackage{csquotes}
\usepackage[inline]{enumitem}
\usepackage{mathtools}
\usepackage{nth}
\usepackage{circledsteps}
\usepackage{balance} %

\newboolean{showcomments}
\setboolean{showcomments}{false}

\newcommand\important[1]{\todo[inline]{\textbf{Important:} #1}}
\newcommand\leo[1]{\todo[color=orange,inline]{\textbf{Leo:} #1}}
\newcommand\michal[1]{\todo[color=green,inline]{\textbf{Michał:} #1}}
\newcommand\onur[1]{\todo[color=red,inline]{\textbf{Onur:} #1}}
\newcommand\navin[1]{\todo[color=brown,inline]{\textbf{Navin:} #1}}
\newcommand\maciej[1]{\todo[color=violet,inline]{\textbf{Maciej:} #1}}
\newcommand\bjorn[1]{\todo[color=yellow,inline]{\textbf{Björn:} #1}}
\ifthenelse{\boolean{showcomments}} { }
{
\renewcommand\important[1]{}
\renewcommand\leo[1]{}
\renewcommand\michal[1]{}
\renewcommand\onur[1]{}
\renewcommand\navin[1]{}
\renewcommand\maciej[1]{}
\renewcommand\bjorn[1]{}
}

\ifthenelse{\boolean{showcomments}}
{ \newcommand{\mynote}[3]{
    \protect\fbox{\bfseries\sffamily\scriptsize#1}
    {\small\textsf{\emph{\color{#3}{#2}}}}}}
{ \newcommand{\mynote}[3]{}}

\newboolean{enablevspacehax}
\setboolean{enablevspacehax}{false}
\ifthenelse{\boolean{enablevspacehax}}
{\newcommand{\vspacehax}[1]{\vspace{#1}}}
{\newcommand{\vspacehax}[1]{}}

\newcommand{\cf}{cf.\@\xspace}

\newcommand{\etal}{\textit{et al.}\@\xspace}
\newcommand{\eg}{\textit{e.g.}\@\xspace}

\newcommand{\ie}{\textit{i.e.}\@\xspace}

\definecolor{verylightgray}{gray}{0.8}

\newcolumntype{L}{l<{\hspace{1cm}}}
\newcolumntype{C}{c<{\hspace{1cm}}}
\newcolumntype{D}{c<{\hspace{0.3cm}}}

\crefname{section}{Sec.}{Sec.}
\Crefname{section}{Section}{Sections}
\crefname{equation}{eq.}{eq.}
\crefname{figure}{Fig.}{Fig.s}
\Crefname{figure}{Figure}{Figures}

\floatstyle{plain}
\newfloat{lstfloat}{htbp}{lop}
\floatname{lstfloat}{Listing}
\crefalias{lstfloat}{listing}

\newcommand{\bs}{Bitswap\xspace}

\newcommand{\mtputprov}{\textsc{PutProvider}}
\newcommand{\mtfn}{\textsc{FindNode}}

\newcommand{\dhtGetClosestPeers}[1]{\ensuremath{\Call{GetClosestPeers}{#1}}}
\newcommand{\dhtProvide}[1]{\ensuremath{\Call{Provide}{#1}}}
\newcommand{\dhtFindProviders}[1]{\ensuremath{\Call{FindProviders}{#1}}}

\newcommand{\node}{\ensuremath{v}\xspace}
\newcommand{\key}{\textit{key}\xspace}

\newcommand{\dataitem}{\ensuremath{d}\xspace}
\newcommand{\cid}{\ensuremath{c}\xspace}

\newcommand{\dhtgraph}{\ensuremath{G_\text{DHT}}}

\lstdefinelanguage{Golang}%
{morekeywords=[1]{package,import,func,type,struct,return,defer,panic,%
		recover,select,var,const,iota,},%
	morekeywords=[2]{string,uint,uint8,uint16,uint32,uint64,int,int8,int16,%
		int32,int64,bool,float32,float64,complex64,complex128,byte,rune,uintptr,%
		error,interface},%
	morekeywords=[3]{map,slice,make,new,nil,len,cap,copy,close,true,false,%
		delete,append,real,imag,complex,chan,},%
	morekeywords=[4]{for,break,continue,range,goto,switch,case,fallthrough,if,%
		else,default,},%
	morekeywords=[5]{Println,Printf,Error,},%
	sensitive=true,%
	morecomment=[l]{//},%
	morecomment=[s]{/*}{*/},%
	morestring=[b]',%
	morestring=[b]",%
	morestring=[s]{`}{`},%
}

\lstdefinelanguage{JavaScript}{
	keywords={typeof, new, true, false, catch, function, return, null, catch, switch, var, if, in, while, do, else, case, break},
	keywordstyle=\color{blue}\bfseries,
	ndkeywords={class, export, boolean, throw, implements, import, this},
	ndkeywordstyle=\color{darkgray}\bfseries,
	identifierstyle=\color{black},
	sensitive=false,
	comment=[l]{//},
	morecomment=[s]{/*}{*/},
	morestring=[b]',
	morestring=[b]"
}

\lstdefinelanguage{Protobuf}{
	keywords={message, option, repeated, bytes, string, uint64, enum, package, syntax, required},
	keywordstyle=\color{blue}\bfseries,
	ndkeywords={},
	ndkeywordstyle=\color{darkgray}\bfseries,
	identifierstyle=\color{black},
	sensitive=false,
	comment=[l]{//},
	morecomment=[s]{/*}{*/},
	morestring=[b]',
	morestring=[b]"
}

\usetikzlibrary{positioning}
\usetikzlibrary{graphs}
\usetikzlibrary{shapes.multipart}
\usetikzlibrary{shapes.misc}
\usetikzlibrary{matrix}
\usetikzlibrary{shapes}
\usetikzlibrary{arrows}
\usetikzlibrary{fit}
\usetikzlibrary{calc}
\usetikzlibrary{decorations.pathreplacing}

\tikzstyle{startstop} = [rectangle, rounded corners, minimum width=3cm, minimum height=1cm,text centered, draw=black, fill=red!30]
\tikzstyle{io} = [trapezium, trapezium left angle=70, trapezium right angle=110, minimum width=2.5cm, text width=2.5cm, minimum height=1cm, text centered, draw=black, fill=blue!30]
\tikzstyle{process} = [rectangle, minimum width=2.5cm, minimum height=1cm, text width=2.5cm, text centered, draw=black, fill=orange!30]
\tikzstyle{decision} = [diamond, aspect=3, minimum width=3cm, minimum height=1cm, text centered, draw=black, fill=green!30]
\tikzstyle{arrow} = [thick,->,>=stealth]
\tikzstyle{darrow} = [thick,<->,>=stealth]
\tikzstyle{entity_own} = [rectangle, minimum width=2.5cm, minimum height=1cm, text width=2.5cm, text centered, draw=black, fill=orange!30]
\tikzstyle{entity_foreign} = [rectangle, minimum width=2.5cm, minimum height=1cm, text width=2.5cm, text centered, draw=black, fill=gray!10]
\tikzstyle{functionality} = [rectangle, minimum width=2.5cm, minimum height=1cm, text width=2.5cm, text centered, draw=black, fill=white]
\tikzstyle{dataset} = [rectangle, minimum width=2.5cm, minimum height=1cm, text width=2.5cm, text centered, draw=black, fill=blue!10]
\tikzstyle{circlenode} = [circle, inner sep=0.05cm, draw=black, fill=gray!10, text centered]
\tikzstyle{ordernode} = [circle, thick, inner sep=0.02cm, draw=black, text centered]

\copyrightyear{2023}
\acmYear{2023}
\setcopyright{acmlicensed}
\acmConference[IMC '23]{Proceedings of the 2023 ACM Internet Measurement Conference}{October 24--26, 2023}{Montreal, QC, Canada}
\acmBooktitle{Proceedings of the 2023 ACM Internet Measurement Conference (IMC '23), October 24--26, 2023, Montreal, QC, Canada}
\acmPrice{15.00}
\acmDOI{10.1145/3618257.3624797}
\acmISBN{979-8-4007-0382-9/23/10}



\begin{document}

\begin{acronym}[Derp]
	\acro{p2p}[P2P]{peer-to-peer}
	\acro{dht}[DHT]{distributed hash table}
	\acro{ipfs}[IPFS]{Interplanetary Filesystem}
	\acro{ipld}[IPLD]{Interplanetary Linked Data}
	\acro{json}[JSON]{JavaScript Object Notation}
	\acro{cbor}[CBOR]{Concise Binary Object Notation}
	\acro{sfs}[SFS]{Self-Certifying Filesystem}
	\acro{ipns}[IPNS]{Interplanetary Namesystem}
	\acro{cid}[CID]{content identifier}
	\acro{dag}[DAG]{directed acyclic graph}
	\acro{pb}[Protobuf]{Protocl Buffers}
	\acro{cdf}[CDF]{cumulative distribution function}
	\acro{pdf}[PDF]{probability density function}
	\acro{nft}[NFT]{non-fungible token}
	\acro{url}[URL]{uniform resource locator}
	\acro{uri}[URI]{uniform resource identifier}
	\acro{ccn}[CCN]{content-centric networking}
	\acro{crdt}[CRDT]{concurrent(?) replicated data type}
	\acro{http}[HTTP]{Hypertext Transfer Protocol}
	\acro{api}[API]{application programming interface}
	\acro{csv}[CSV]{comma-separated values}
	\acro{quic}[QUIC]{QUIC}
	\acro{tcp}[TCP]{Transmission Control Protocol}
	\acro{ipns}[IPNS]{InterPlanetary Name System}
	\acro{ens}[ENS]{Ethereum Name System}
\end{acronym}

\title{The Cloud Strikes Back: Investigating the Decentralization of IPFS}

\author{Leonhard Balduf}
\orcid{0000-0002-3519-7160}
\affiliation{%
	\institution{Technical University of Darmstadt}
	\city{Darmstadt}
	\country{Germany}}
\email{leonhard.balduf@tu-darmstadt.de}

\author{Maciej Korczyński}
\orcid{0000-0002-4334-3260}
\affiliation{%
    \institution{Univ. Grenoble Alpes} %
    \city{Grenoble}
	\country{France}
	}
\email{maciej.korczynski@univ-grenoble-alpes.fr}

\author{Onur Ascigil}
\orcid{0000-0002-3023-6431}
\affiliation{%
	\institution{Lancaster University}
	\city{Lancaster}
	\country{United Kingdom}}
\email{o.ascigil@lancaster.ac.uk}

\author{Navin V. Keizer}
\orcid{0000-0002-2417-3390}
\affiliation{%
	\institution{University College London}
	\city{London}
	\country{United Kingdom}}
\email{navin.keizer.15@ucl.ac.uk}

\author{George Pavlou}
\orcid{0000-0001-8211-2740}
\affiliation{%
	\institution{University College London}
	\city{London}
	\country{United Kingdom}}
\email{g.pavlou@ucl.ac.uk}

\author{Björn Scheuermann}
\orcid{0000-0002-1133-1775}
\affiliation{%
	\institution{Technical University of Darmstadt}
	\city{Darmstadt}
	\country{Germany}}
\email{scheuermann@kom.tu-darmstadt.de}

\author{Michał Król}
\orcid{0000-0002-3437-8621}
\affiliation{%
	\institution{City, University of London}
	\city{London}
	\country{United Kingdom}}
\email{michal.krol@city.ac.uk}

\begin{abstract}
Interplanetary Filesystem (IPFS) is one of the largest peer-to-peer filesystems in operation.
The network is the default storage layer for Web3 and is being presented as a solution to the centralization of the web.
In this paper, we present a large-scale, multi-modal measurement study of the IPFS network.
We analyze the topology, the traffic, the content providers and the entry points from the classical Internet.

Our measurements show significant centralization in the IPFS network and a high share of nodes hosted in the cloud. 
We also shed light on the main stakeholders in the ecosystem. 
We discuss key challenges that might disrupt continuing efforts to decentralize the Web and highlight multiple properties that are creating pressures toward centralization.
\end{abstract}

\maketitle
\acresetall

\leo{From the IMC Website: All papers must include, in a clearly marked appendix section with the heading “Ethics”, a statement about ethical issues; papers that do not include such a statement may be rejected. This could be, if appropriate for the paper, simply the sentence “This work does not raise any ethical issues.”. If the work involves human subjects or potentially sensitive data (e.g., user traffic or social network information, evaluation of censorship, etc.), the paper should clearly discuss these issues, perhaps in a separate subsection.}
\leo{Also: You should ensure that the paper prints well on black-and-white printers, not color printers. Pay particular attention to figures and graphs in the paper to ensure that they are legible without color. Explicitly using grayscale colors will provide best control over how graphs and figures will print on black-and-white printers.}

\section{Introduction}\label{sec:intro}

\ac{ipfs}~\cite{benet2014ipfs} is one of the largest \ac{p2p} filesystem currently in operation.
The platform underpins various decentralized web applications~\cite{ecosystem},
including social networking and discussion~\cite{discussify, matters},
data storage~\cite{Space, Peergos, temporal},
content search~\cite{almonit, deece},
messaging~\cite{Berty},
content streaming~\cite{Audius, dtube, Watchit}
gaming~\cite{gala, Splinterlands},
and e-commerce~\cite{ethlance, dClimate}.
\ac{ipfs} is widely used as external storage for blockchain-based applications, including valuable NFT platforms~\cite{balduf2022nft,das2022nft}. 

The IPFS network currently contains a steady number of $\approx 30,000$~\cite{dashboard} online nodes, spread across 2,700 Autonomous Systems and 152 countries, according to a recent study~\cite{trautwein2022design} that also observed widespread usage by clients with 7.1 million content retrieval operations observed from a single vantage point and during a single day.
Support for accessing IPFS has further been integrated into HTTP gateways (\eg, Cloudflare) and mainstream browsers such as Opera and Brave, allowing easy uptake.

IPFS is being presented as the default storage layer for Web3 with a strong focus on decentralization~\cite{ipfs_website}.
Storage decentralization is supposed to offer multiple benefits~\cite{ramachandran2020towards, hasan2005survey}.
Data is spread among many replicas, making privacy-intrusive data mining more difficult.
Data ownership is more transparent, and the lack of centralization makes the overall system more robust against technical, legal or regulatory attacks.
However, these properties may also bring inherent challenges that are difficult to avoid, particularly when considering the natural pressures towards centralization in both social~\cite{wilson2009user} and economic~\cite{stigler1958economies} systems.

\paragraph{Contributions}
In this paper, we evaluate the current state of the IPFS network with a focus on decentralization and make the following contributions.
We build tools for multidimensional observation of the IPFS network.
In contrast to previous studies~\cite{trautwein2022design}, we not only discover the system participants but also the traffic generated by the network.
This includes the \ac{dht}~\cite{maymounkov2002kademlia} and \bs{}~\cite{de2021accelerating}, the two core IPFS protocols used for data discovery and exchange.
Furthermore, we observe multiple entry points to the IPFS ecosystems (\eg, HTTP gateways, browser extensions, \ac{ens}) using passive and active DNS measurements.
We then analyze our 9-month dataset to provide insights into the state of the network, content exchange patterns and peer behaviour.
We assess the centralization of the network and its reliance on cloud components.
Finally, we discuss the drivers behind centralization and explore techniques that could reduce this propensity.

\paragraph{Findings}
Overall, our main findings include
\begin{enumerate*}
	\item We observe that almost 80\% of the IPFS DHT servers are hosted in the cloud with the top 3 cloud providers hosting %
$51.9$\%%
 of the servers.
	Our results paint a different picture of the IPFS network from the one presented in a recent study~\cite{trautwein2022design} reporting less than 3\% of the cloud-based nodes. We explain the reason behind the differences and show how small changes in the measurement methodology can lead to different conclusions.
	\item We found that the network experiences a high degree of traffic centralization. The top $5\%$ of the nodes are responsible for up to $95\%$ of the traffic with the largest cloud provider, Amazon-AWS, generating $96\%$ of all the content resolution requests. We also found cloud-based storage platforms such as \emph{nft-storage} or \emph{web3-storage} holding a major share of persistent content in the network.	
	\item We show that content storage is heavily reliant on the cloud infrastructure. Nearly $95\%$ of the content is provided by at least one cloud-based node. Furthermore, many non-cloud providers use cloud nodes as proxies for NAT traversal.
	\item We show that major CDN players, such as Cloudflare, dominate the IPFS HTTP gateway ecosystem. Furthermore, even IPFS content referenced by the decentralized Ethereum Name System (ENS) is mostly stored by a handful of major cloud providers.
\end{enumerate*}

\section{Background}\label{sec:background}
In this section, we provide the necessary background information to understand the systems involved, and the methodology used to produce the measurements presented in this work.
We start with a description of \ac{ipfs} in general,
followed by detailed explanations of its network protocols, content provision and retrieval mechanisms,
as well as DNSLink and \ac{http} gateway functionality.

\paragraph{IPFS}
\label{sec:background:ipfs}
IPFS is a content-centric network where nodes are identified via their \emph{peer ID}, which is derived from the public key of a unique key pair.
By default, nodes maintain the same ID over time but a new one can be generated on request.
Each node advertises a set of network endpoints describing their IP address, transport protocol and port number.
One peer ID can be associated with multiple endpoints (\eg, multihoming),
and one IP address can be associated with multiple peer IDs (\eg, when hosting multiple nodes on a single machine).

In \ac{ipfs}, each piece of content is identified by a \ac{cid}.
A \ac{cid} for item \dataitem{} is derived by hashing the content of \dataitem{}, so that $\text{CID}(\dataitem) = h(\dataitem)$ for some cryptographic hash function $h$.\footnote{In practice, the CIDs include some metadata and are encoded using a self-describing format.
	We refer the readers to other studies~\cite{trautwein2022design} for a more detailed description.}

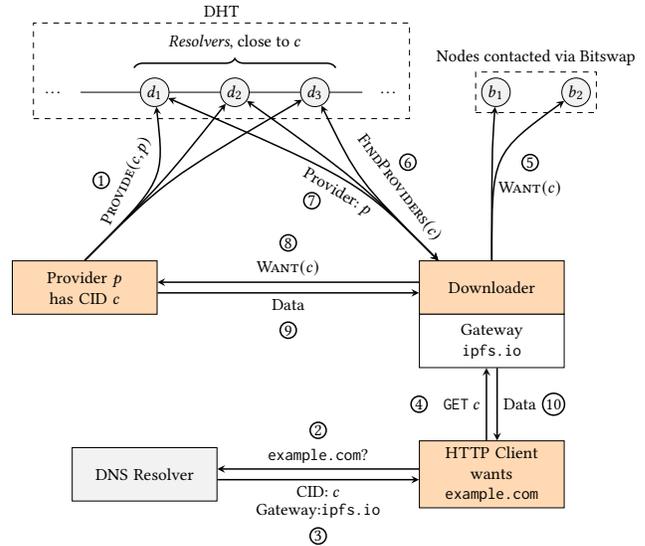
\begin{figure}
	\centering
	\resizebox{\linewidth}{!}{
		\begin{tikzpicture}[node distance=1.5cm]
	\node (dht-line-left) {};
	\node[right of=dht-line-left, xshift=4cm] (dht-line-right) {};
	\draw (dht-line-left) -- (dht-line-right);
	\node[anchor=east] (dht-dots-left) at (dht-line-left.west) {\dots};
	\node[anchor=west] (dht-dots-right) at (dht-line-right.east) {\dots};
	
	\node[circlenode, right of=dht-line-left] (d1) {$d_1$};
	\node[circlenode, right of=d1] (d2) {$d_2$};
	\node[circlenode, right of=d2] (d3) {$d_3$};

	\node[fit=(d1) (d2) (d3)] (close-nodes-fit) {};
	
	\draw [thick,decorate,decoration={brace,amplitude=5pt,raise=0.5ex}]
	(close-nodes-fit.north west) -- (close-nodes-fit.north east);
	\node[anchor=south, yshift=1em] at (close-nodes-fit.north) (closeness-annotation) {\emph{Resolvers}, close to $c$};

	\node[draw=black, dashed, fit=(dht-dots-left) (d1) (d2) (d3) (dht-dots-right) (close-nodes-fit) (closeness-annotation)] (networkfit) {};
	\node[above] at (networkfit.north) {DHT};

	\node[circlenode, right of=dht-dots-right, xshift=.5cm] (b1) {$b_1$};
	\node[circlenode, right of=b1] (b2) {$b_2$};

	\node[draw=black, dashed, fit=(b1) (b2)] (bsfit) {};
	\node[above] at (bsfit.north) {Nodes contacted via Bitswap};

	\node[entity_own, below left of=networkfit, yshift=-3cm, xshift=-1.5cm] (provider) {Provider $p$\\has CID $c$};
	\node[entity_own, below right of=networkfit, yshift=-3cm, xshift=4cm] (downloader) {Downloader};
	\node[functionality, below of=downloader,yshift=0.5cm] (gateway-func) {Gateway\\\texttt{ipfs.io}};
	\node[entity_own, below of=gateway-func,yshift=-1cm] (http-client) {HTTP Client\\wants \texttt{example.com}};
	\node[entity_foreign, left of=http-client, xshift=-5cm] (dnslink) {DNS Resolver};

	\draw[arrow] ([yshift=0.1cm]http-client.west) -- ([yshift=0.1cm]dnslink.east) node[midway, above] (l2) {\texttt{example.com}?};
	\draw[arrow] ([yshift=-0.1cm]dnslink.east) -- ([yshift=-0.1cm]http-client.west) node[midway, below] {CID: $c$} node[midway, below, yshift=-1em] (l3) {Gateway:\texttt{ipfs.io}};

	\draw[arrow] ([xshift=-0.1cm]http-client.north) -- ([xshift=-0.1cm]gateway-func.south) node[midway, left] (l4) {\texttt{GET} $c$};
	\draw[arrow] ([xshift=0.1cm]gateway-func.south) -- ([xshift=0.1cm]http-client.north) node[midway, right] (l10) {Data};

	\draw[arrow] (provider.north) .. controls ([yshift=1.5cm,xshift=1.5cm]provider.north) .. (d1) node [midway, above, sloped, xshift=-0.2cm, yshift=0.2cm] (l1) {\textsc{Provide}$(c,p)$};
	\draw[arrow] (provider.north) .. controls ([yshift=1.5cm,xshift=1.5cm]provider.north) .. (d2);
	\draw[arrow] (provider.north) .. controls ([yshift=1.5cm,xshift=1.5cm]provider.north) .. (d3);

	\draw[darrow] ([xshift=-1cm]downloader.north) .. controls ([yshift=1.5cm,xshift=-2.5cm]downloader.north) .. (d1)
	node [midway, below, sloped] (l7) {Provider: $p$};
	\draw[darrow] ([xshift=-1cm]downloader.north) .. controls ([yshift=1.5cm,xshift=-2.5cm]downloader.north) .. (d2);
	\draw[darrow] ([xshift=-1cm]downloader.north) .. controls ([yshift=1.5cm,xshift=-2.5cm]downloader.north) .. (d3)
	node [midway, above, sloped,xshift=.5cm,yshift=.2cm] (l6) {\textsc{FindProviders}$(c)$};

	\draw[arrow] (downloader.north) .. controls ([yshift=2cm]downloader.north) .. (b1) node [midway, right, yshift=-0.5cm] (l5) {\textsc{Want}$(c)$};
	\draw[arrow] (downloader.north) .. controls ([yshift=2cm]downloader.north) .. (b2);

	\draw[arrow] ([yshift=0.1cm]downloader.west) -- ([yshift=0.1cm]provider.east)
	node [midway, above] (l8) {\textsc{Want}$(c)$};
	\draw[arrow] ([yshift=-0.1cm]provider.east) -- ([yshift=-0.1cm]downloader.west)
	node [midway, below] (l9) {Data};

	\node[ordernode] at ([xshift=-1.5em, yshift=0cm]l1) {1};
	\node[ordernode] at ([xshift=0cm, yshift=1.5em]l2) {2};
	\node[ordernode] at ([xshift=0cm, yshift=-1.5em]l3) {3};
	\node[ordernode] at ([xshift=-0.8cm, yshift=0cm]l4) {4};
	\node[ordernode] at ([xshift=0cm, yshift=1.5em]l5) {5};
	\node[ordernode] at ([xshift=0.5em, yshift=1.5em]l6) {6};
	\node[ordernode] at ([xshift=-1.5em, yshift=-0.5em]l7) {7};
	\node[ordernode] at ([xshift=0cm, yshift=1.5em]l8) {8};
	\node[ordernode] at ([xshift=0cm, yshift=-1.5em]l9) {9};
	\node[ordernode] at ([xshift=2em, yshift=0cm]l10) {10};
\end{tikzpicture}}\unskip%
	\vspacehax{-0.3cm}%
	\caption[Illustrations of IPFS Provide and Request Functionalities.]{
		Illustrations of IPFS Provide and Request Functionalities.
	}%
	\label{fig:dht_diagrams}
    \vspacehax{-0.5cm}%
\end{figure}

\acp{cid} do not contain information about the content location, are immutable (changing the content generates a new CID), and are not human-readable.
This enables easy content deduplication, data retrieval from the closest available location, and maintaining data integrity.
However, it also means that a downloader first needs to resolve a CID to a list of providers, \ie, nodes storing the content, before the actual content transfer.
The resolution is done using a Kademlia~\cite{maymounkov2002kademlia} \ac{dht} and the \bs{} protocol.
\Cref{fig:dht_diagrams} shows typical interactions between entities in the network, which we will now explain in more detail.

\paragraph{Bitswap}
\label{sec:background:bitswap}
\bs~\cite{de2021accelerating} is a simple protocol used to exchange blocks of data.
Typically, IPFS nodes maintain \bs connections to a few hundred random peers.\footnote{This number may differ depending on the configuration.}
The protocol allows one to ask a peer whether it has a target block or to directly request and transfer the target content (\cf \Cref{fig:dht_diagrams} \Circled{5}).

\paragraph{DHT}
\label{sec:background:dht}
\ac{ipfs} uses a Kademlia~\cite{maymounkov2002kademlia} \ac{dht} implementing a key-value store.
A new participant node joins the \ac{ipfs} network by contacting one of the hardcoded bootstrap nodes.
This bootstrap node provides the new node with some initial peers allowing it to join the \ac{dht}.
The new node uses this information to perform a walk through the \ac{dht} towards its own peer ID to discover peers and fill its routing table. 

The main operation \dhtGetClosestPeers{\key} traverses the DHT and returns the $k$ closest peers to the target \key{}.
In each step, the querying node contacts the closest nodes to \key it knows of.
Each of these peers returns the $k$ closest peers to \key{} in its own routing table and the addresses of these peers.
The querying node again sends requests to the peers closest to \key{}, among peers it just received.
This process repeats until the client does not find any more peers closer to \key{}.

Recent versions of the software differentiate between \ac{dht} \emph{servers} and \ac{dht} \emph{clients}.
The latter only use the \ac{dht} as a service for resolution and routing, which is provided through \ac{dht} servers.
To become a \ac{dht} server, the software determines whether it is connectable from the internet (as opposed to, \eg, NAT-ed).
Only connectable nodes become \ac{dht} servers unless the user explicitly modifies their configuration.
Generally, the set of \ac{dht} clients can be understood as the user-operated fringe of the network, consisting of nodes behind NAT, whereas the \ac{dht} servers form the network's core.

\paragraph{Content Advertisement}
When a user adds content to the network, it adds it to its local node and uses the DHT to advertise itself as a \emph{provider} for the CID representing the content.
First, it creates a \emph{provider record} that contains \cid and its own network information.
During a \dhtProvide{\cid} operation, the provider first uses \dhtGetClosestPeers{\cid} to locate the $k=20$ peers closest to \cid{},
and then sends them a \mtputprov{} message including the provider record.
We call the peers that hold provider records for \cid{} the \emph{resolvers} for \cid{}.
(\cf \Cref{fig:dht_diagrams} \Circled{1})

By default, each IPFS client becomes a provider for each piece of content it downloads and automatically registers itself as a provider for the corresponding CID.
As a result, the system provides an auto-scaling feature with supply automatically rising with demand.

A provider without a public IP address (\eg DHT client) cannot directly receive download requests for the content it provides, unless it is already connected to the downloader via \bs{}.
Generally, NAT-ed nodes first establish a connection to a random DHT server supporting the relay protocol that will act as a reverse proxy and NAT-punching introducer.
The provider includes the IP address of the proxy in the provider records it generates.
As of v0.13, \ac{ipfs} includes a NAT hole-punching mechanism called direct connection upgrade through a relay (DCUtR),
which is functionally similar to the one used in the SSU protocol of the I2P network~\cite{dcutr,i2pssu}.

\paragraph{Content Retrieval}
\label{sec:background:content_retrieval}
Downloading a data item \dataitem{} with \ac{cid} \cid{} is a two-step process:
\begin{enumerate*}
	\item providers for \cid{} are found (\cf \Cref{fig:dht_diagrams} \Circled{5}, \Circled{6}, \Circled{7}),
	\item connections to the providers are established and \dataitem{} is downloaded from them directly via \bs{} (\Circled{8}, \Circled{9}).
\end{enumerate*}

The search for providers begins with a local, 1-hop broadcast via \bs{} (\Circled{5}) to all connected neighbours looking for the target \ac{cid}.
Searching via \bs{} is fast, but does not provide reliable content resolution, in particular for less popular or new content.
If this does not yield any results, the downloader invokes \dhtFindProviders{\cid} using the \ac{dht} (\Circled{6}).
This operation uses a \ac{dht} walk identical to that of \dhtGetClosestPeers{\cid} to find up to $k$ resolvers but also queries encountered nodes for a provider record for \cid{}.
The process terminates when either $20$ providers have been found, or all resolvers have been asked (\Circled{7}).
The downloader concurrently initiates \bs{} connections to the discovered providers and retrieves the requested content (\Circled{8}, \Circled{9}).
\paragraph{Entry Points}
Accessing content with IPFS can be considered complex due to a few reasons.
Downloading IPFS content requires installing additional software, joining the \ac{p2p} network and using custom protocols.
While \acp{cid} ensure content integrity and prevent tampering, they can be long and difficult for users to remember or share compared to traditional URLs or domain names.
Furthermore, the identifier changes with every modification of the content.
For instance, modifying a website hosted on IPFS creates a completely new CID that must be communicated to all the website viewers.
Another limitation is browser support: traditional web browsers are designed to work with the HTTP/HTTPS protocols and may not have native support for \ac{ipfs}.
To simplify content access, IPFS implements multiple tools bridging the gap with the traditional Web.%

\paragraph{HTTP Gateways}
\label{sec:background:gateways}
Gateways translate \ac{http} \textsc{GET} requests to content retrievals in \ac{ipfs} and enable IPFS-agnostic users to access the content (\cf \Cref{fig:dht_diagrams} \Circled{4}, \Circled{10}).

When a gateway receives an \ac{http} \textsc{GET} for a CID, it 
\begin{enumerate*}
	\item checks its local cache
	\item finds and downloads the content using IPFS, and
	\item returns the content to the client using \ac{http}.
\end{enumerate*}
Protocol Labs maintains a list of public gateways~\cite{gateway_checker}, some of which are operated by large content delivery networks such as Cloudflare.
Previous studies showed extensive usage of gateways and their noticeable share of traffic in the IPFS network~\cite{balduf2022nft, balduf2022monitoring}.

\paragraph{DNSLink}
\label{sec:background:dnslink}
DNSLink~\cite{dnslink} enables content publishers to associate domain names with IPFS content.
It integrates the traditional DNS with IPFS, enabling users to access IPFS content using familiar domain names instead of intricate \acp{cid} (\cf \Cref{fig:dht_diagrams} \Circled{2}, \Circled{3}).

DNSLink leverages DNS records to establish a connection between a specified domain name, such as \texttt{example.com}, and an IPFS address.
It is achieved by storing a DNS \texttt{TXT} record within a dedicated subdomain beginning with the \texttt{\_dnslink} label (e.g., \texttt{\_dnslink.example.com}).
The structure of the \texttt{TXT} record follows the guidelines outlined in RFC 1464~\cite{rfc1464}, which defines a formatted representation as \texttt{<key>=<value>}.
Within the DNS \texttt{TXT} record, one can find either of the following entries: \texttt{dnslink=/ipfs/<CID>} or \texttt{dnslink=/ipns/<hash of public key>}.
The first one associates the \ac{cid} directly with the domain name, whereas the second one associates the IPNS key's hash value with the domain.
The second approach enables redirecting users, e.g., a website visitor, to the most recent version of an object in IPFS, considering that modifying an object alters its \ac{cid}.

\begin{sloppypar}
To ensure that content stored on IPFS can be accessed, domain name owners need to configure their root domain (e.g., \texttt{example.com}) or subdomain (e.g., \texttt{subdomain.example.com}) to point to an IPFS gateway or proxy server.
This configuration can be done in two ways:
\begin{enumerate*}
	\item assigning the IP address of the IPFS gateway or proxy server as the value of the \texttt{A} record for the domain, or
	\item setting a \texttt{CNAME} or \texttt{ALIAS} record that matches the domain of the IPFS gateway or proxy server.
\end{enumerate*}
By following either of these methods, domain name owners can establish the necessary connection between their domain or subdomain and the IPFS gateway or proxy server, enabling the retrieval of content stored on IPFS.
\end{sloppypar}

If the DNS provider supports \texttt{ALIAS} records, they are generally recommended for pointing the root domain to an IPFS gateway or proxy server.
For instance, to configure the \texttt{example.com} domain, a domain name owner can add the following \texttt{ALIAS} record in the zone file:
\texttt{example.com ALIAS gateway.ipfs.io}, directing \texttt{example.com} to a public gateway operated by Protocol Labs.
Similarly, a subdomain can be configured with a \texttt{CNAME} record such as \texttt{subdomain.example .com CNAME cloudflare-ipfs.com},
directing it to a public gateway operated by Cloudflare.

\paragraph{Ethereum Name Service (ENS)}\label{sec:background:ens}
The Ethereum Name Service (ENS) \cite{ens} is an alternative name-registry service for Web3 which allows users to register name-value pairs directly on the Ethereum blockchain~\cite{wood2014ethereum}. One of the prominent use cases of ENS is to provide a mapping from human-readable domain names to cryptographic hashes such as IPFS CIDs, without relying on the centralization in the current DNS infrastructure in the form of Top-Level Domain (TLD) ownership and reliance on ICANN.%

Namespace management in ENS is governed by several smart contracts. The \textit{Registry} maintains a top-level mapping of all domains and subdomains to their owner, resolver, and caching time-to-live. \textit{Registrar} contracts maintain ownership of individual domains (\eg \texttt{.eth}) and their subdomains. Finally, the \textit{resolver} contract for a (sub)domain points towards a value mapping set by the owner such as an Ethereum address or IPFS CID and assists users in resolving names in a decentralized manner.

\section{Methodology and Collected Datasets}
\label{sec:methodology_datasets}

\begin{figure*}[ht]
	\centering%
	\scalebox{0.8}{
		\begin{tikzpicture}[node distance=3cm]
	\node (name3) {IPFS Node};
	\node[functionality, below of=name3,yshift=2cm] (provider3)  {Content Provider Functionality};
	\node[functionality, below of=provider3,yshift=1cm] (requester3) {Content Requester Functionality};
	\node[functionality, below of=requester3] (gateway3) {(Public) Gateway Functionality};
	\node[functionality, below of=gateway3,yshift=1.5cm] (buckets3) {DHT Buckets};

	\node[entity_foreign, fit=(name3) (gateway3) (provider3) (requester3) (buckets3)] (ipfs-node2) {};

	\node[xshift=-0.1cm, yshift=0.1cm] (name2) {IPFS Node};
	\node[functionality, below of=name2, yshift=2cm] (provider2)  {Content Provider Functionality};
	\node[functionality, below of=provider2,yshift=1cm] (requester2) {Content Requester Functionality};
	\node[functionality, below of=requester2] (gateway2) {(Public) Gateway Functionality};
	\node[functionality, below of=gateway2 ,yshift=1.5cm] (buckets2) {DHT Buckets};

	\node[entity_foreign, fit=(name2) (gateway2) (provider2) (requester2) (buckets2)] (ipfs-node) {};

	\node[below] at (ipfs-node2.south) (ipfs-node-dots) {\vdots};

	\node[xshift=-0.1cm, yshift=0.1cm] (name) {IPFS Node};
	\node[functionality, below of=name, yshift=2cm] (provider) {Content Provider Functionality};
	\node[functionality, below of=provider,yshift=1cm] (requester) {Content Requester Functionality};
	\node[functionality, below of=requester] (gateway) {Public Gateway Functionality};
	\node[functionality, below of=gateway,yshift=1.5cm] (buckets) {DHT Buckets};

	\draw[arrow] ([xshift=-0.1cm]gateway.north) -- ([xshift=-0.1cm]requester.south) node [midway, above, sloped] {Requests};
	\draw[arrow] ([xshift=0.1cm]requester.south) -- ([xshift=0.1cm]gateway.north) node [midway, above, sloped] {Data};
	\draw[darrow] (provider) -- (requester) node [midway, right] (tn-caching) {Caching};

	\node[draw=black, dashed, fit=(ipfs-node2) (ipfs-node) (ipfs-node-dots) (tn-caching)] (networkfit) {};
	\node[above] at (networkfit.north) {IPFS Network};

	\node[entity_own, right of=requester, xshift=2cm, yshift=-1.75cm] (monitor) {Bitswap Monitoring Node};
	\node[functionality, above of=monitor, yshift=-1.5cm] (aggregate) {Aggregate Daily};
	\node[functionality, right of=aggregate] (dedup) {Extract CIDs,\\Deduplicate};
	\node[functionality, above of=dedup,yshift=-1.5cm] (sample) {Uniform Sample};
	\node[dataset, left of=sample] (bs-cids-dataset) {Daily Sampled Bitswap CIDs};
	\node[dataset, right of=monitor] (bs-full-dataset) {Raw Bitswap Requests};
	\draw[arrow] (requester) -- (monitor.west) node [midway, above, sloped] {Requests} node [midway, below,sloped] {via Bitswap};
	\draw[arrow] (monitor) -- (aggregate) node [midway, above] (tn) {};
	\draw[arrow] (aggregate) -- (dedup) node [midway, above] (tn2) {};
	\draw[arrow] (dedup) -- (sample) node [midway, above] {};
	\draw[arrow] (sample) -- (bs-cids-dataset);
	\draw[arrow] (monitor) -- (bs-full-dataset);

	\node[draw=black, dashed, fit=(monitor) (aggregate) (sample) (tn) (tn2) (bs-cids-dataset) (bs-full-dataset)] (monitorfit) {};
	\node[above, anchor=south east] at (monitorfit.north east) {Bitswap Monitoring};

	\node[functionality, right of=gateway, xshift=2cm, yshift=-2cm] (req-random-content) {Request};
	\node[functionality, above of=req-random-content, yshift=-1.5cm] (gen-random-content) {Generate Unique Content};
	\node[entity_own, right of=gen-random-content] (gateway-probe) {Gateway Probe};
	\node[dataset, below of=gateway-probe, yshift=1.5cm] (gateway-dataset) {Public Gateway IDs};
	\draw[arrow] (req-random-content) -- (gateway.east) node [midway, above, sloped] {HTTP};
	\draw[arrow] (gateway-probe) -- (gen-random-content) node [midway, above] (tn3) {};
	\draw[arrow] (gen-random-content) -- (req-random-content);
	\draw[arrow] (monitor) -- (gateway-probe) node [midway, right] {Requests};
	\draw[arrow] (gateway-probe) -- (gateway-dataset) node [midway, above] {};
	\draw[arrow] (gen-random-content) -- (monitor) node [midway, left] {Provide via};

	\node[draw=black, dashed, fit=(req-random-content) (gateway-probe) (gateway-dataset) (gen-random-content)] (gateway-probe-fit) {};
	\node[below] at (gateway-probe-fit.south) {Gateway Identification};

	\node[entity_own, above of=bs-cids-dataset, yshift=-1cm] (prov-record-searcher) {Provider Record Searcher};
	\node[dataset, right of=prov-record-searcher] (prov-record-dataset) {Provider Records};

	\draw[arrow] (bs-cids-dataset) -- (prov-record-searcher) node [midway, above] {};
	\draw[arrow] (prov-record-searcher) -- (prov-record-dataset) node [midway, above] {};
	\draw[darrow] (prov-record-searcher.west) -- (provider.east) node [midway, above, sloped] {Find} node [midway, below, sloped] {Providers};

	\node[draw=black, dashed, fit=(prov-record-searcher) (prov-record-dataset)] (prov-record-search-fit) {};
	\node[above] at (prov-record-search-fit.north) {Provider Record Search};

	\node[entity_own, left of=requester, xshift=-1.75cm, yshift=1.5cm] (hydra-booster) {Hydra Booster};
	\node[dataset, above of=hydra-booster, yshift=-1.5cm] (hydra-logs) {Hydra Logs};

	\draw[arrow] (requester.west) -- (hydra-booster.south east) node [midway, above, sloped] {Requests} node [midway, below, sloped] {via DHT};
	\draw[arrow] (provider.west) -- (hydra-booster.east) node [midway, above, sloped] {Provider} node [midway, below, sloped] {Records};
	\draw[arrow] (hydra-booster) -- (hydra-logs);

	\node[draw=black, dashed, fit=(hydra-booster) (hydra-logs)] (hydra-fit) {};
	\node[above] at (hydra-fit.north) {DHT Monitoring};

	\node[entity_own, left of=buckets, xshift=-1.75cm,yshift=-0.5cm] (dht-crawler) {DHT Crawler};
	\node[dataset, above of=dht-crawler, yshift=-1.5cm] (dht-topology) {Overlay Topology};

	\draw[darrow] (dht-crawler) -- (buckets) node [midway, above, sloped] {Extract};
	\draw[arrow] (dht-crawler) -- (dht-topology);

	\node[draw=black, dashed, fit=(dht-crawler) (dht-topology)] (dht-crawling-fit) {};
	\node[below] at (dht-crawling-fit.south) {DHT Crawling};

	\node[entity_own, left of=gateway, xshift=-1.75cm, yshift=1cm] (passive-dns) {Passive DNS Measurements};
	\node[dataset, above of=passive-dns, yshift=-1.5cm] (passive-dns-gateways) {HTTP Gateway Usage};

	\draw[darrow, dashed] (passive-dns.east) -- (gateway.west);
	\draw[arrow] (passive-dns) -- (passive-dns-gateways);

	\node[draw=black, dashed, fit=(passive-dns) (passive-dns-gateways)] (passive-dns-fit) {};
	\node[above] at (passive-dns-fit.north) {Passive DNS};
\end{tikzpicture}}\unskip%
	\vspacehax{-0.3cm}%
	\caption[Data Collection Architecture Overview]{
		Data Collection Architecture, Overview.
	}%
	\label{fig:data_collection_architecture}
    \vspacehax{-0.3cm}%
\end{figure*}
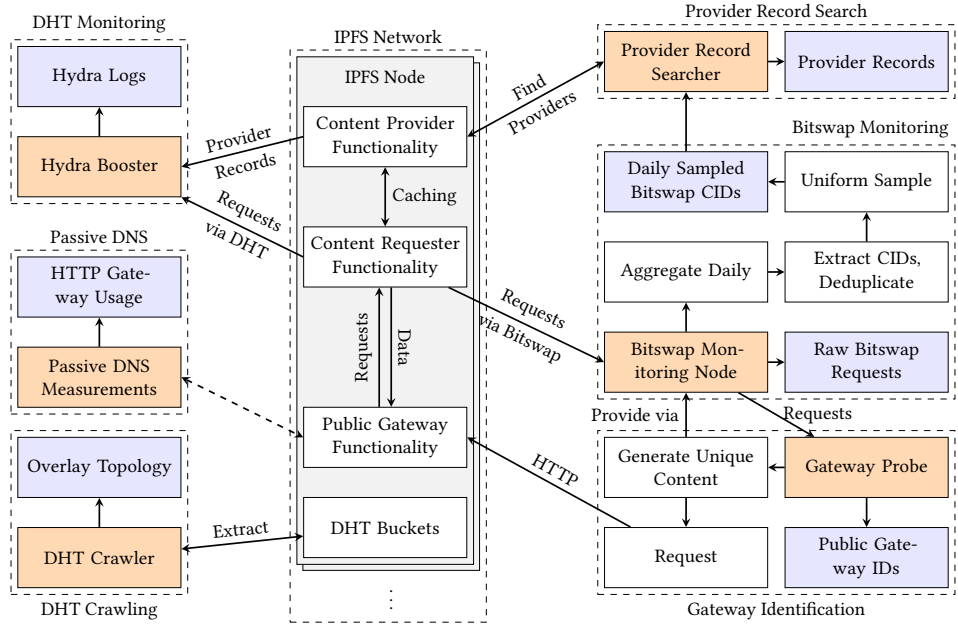

In this section, we explain the methodology employed in this work, and the datasets derived through it.
An architectural overview, showing how different functionalities of \ac{ipfs} nodes are measured, is shown in \Cref{fig:data_collection_architecture}.
We make all code processing data available at~\cite{code_metarepo}.

\paragraph{Topology graph}
As explained in \Cref{sec:background:dht}, \ac{ipfs} builds on top of a Kademlia \ac{dht}, with nodes being partitioned into clients and servers.
A node with address $a_{n}$ stores its outbound \ac{dht}-connections in $k$-buckets, which form a view of the network as a binary trie.
Buckets have a fixed capacity of $k$ connections, which generally leads to the first, furthest, buckets to be filled completely, whereas buckets closer to $a_{n}$ tend to contain fewer and fewer connections.
Only peers providing \ac{dht} server functionality are stored in the buckets.

It is possible to enumerate all \ac{dht} connections of a node through crafted \mtfn{} messages, sweeping the address space towards the target node's own address $a_{target}$.
This process is generally known as \ac{dht} crawling~\cite{henningsen2020mapping,pouwelse2005bittorrent}.

Using the \ac{dht} crawler presented in~\cite{crawler,henningsen2020mapping}, we can enumerate all outgoing \ac{dht} connections of functional \ac{dht} server nodes.
This results in a snapshot of the \ac{dht} graph \dhtgraph{}, at the time of crawling.
In practice, not all nodes are connectable and cooperative, which leads to un-crawlable leaf nodes in the graph.
We crawl the network at least twice per day from %
2023-04-18%
\michal{Minor: could we align the formatting of large numbers? i.e., 20,000 rather than 20 000.}
to %
2023-05-26%
,
for a total of %
$101$%
 snapshots.
We discover an average of %
$25771.6$%
 peers per crawl,
of which %
$17991.4$%
 are connectable and crawlable.
A crawl takes %
$5.0$m%
\michal{remove the ".0" part.} on average, of which the latter half is typically spent waiting on unresponsive peers,
with a connection timeout of $3$m.
Short crawl durations are important to capture accurate snapshots of the network due to churn~\cite{stutzbach2005evaluating,stutzbach2006capturing,falkner2007profiling,daniel2022passively},
and long connection timeouts ensure completeness~\cite{stutzbach2006capturing}.

\paragraph{Counting Methodologies}
We propose a methodology to derive properties of a typical snapshot the \ac{ipfs} \ac{dht}, which is dynamic by default.
Every node on the \ac{ipfs} network is identified by a unique peer ID.
Nodes can announce multiple IP addresses for themselves, which are stored in the DHT.
An example of our dataset, with an additional mapping from IP addresses to geolocation, is shown in \Cref{tab:crawl_data_example}.
From this example, we will try to derive the typical client population and its geospatial distribution.

\begin{table}[ht]
	\centering
	\caption{Example Crawl Dataset}
	\label{tab:crawl_data_example}
	\begin{tabular}{cccc}
		\toprule
		Crawl ID & Peer ID & IP & Geolocation \\
		\midrule
		1 & $p_1$ & $a_1$ & DE \\
		1 & $p_1$ & $a_2$ & DE\\
		1 & $p_2$ & $a_3$ & US\\
		2 & $p_2$ & $a_2$ & DE\\
		2 & $p_2$ & $a_3$ & US\\
		2 & $p_2$ & $a_4$ & US\\
		\bottomrule
	\end{tabular}
\end{table}

Nodes are typically announcing multiple IP addresses, which may differ in geolocation.
The easiest way to resolve this is by counting unique IPs and their mappings, over the entire dataset.
This ignores the time-discrete nature of the network snapshots but gives an estimate of the population of the network over the entire data collection period.
It overcounts peers announcing multiple IPs, especially ones with frequently changing IPs, and includes churning nodes in the count.
For the example dataset, this results in DE=2, US=2.
We refer to this methodology as \emph{Global, Unique IP} (G-IP), which is conceptually similar to the one employed in~\cite{trautwein2022design}.

To combat the problem of overcounting nodes with multiple addresses, we propose to assign each \emph{peer} a single value of the derived property in question, in this case, geolocation.
We can then count peers, the participants of the overlay network, instead of underlay addresses.
We propose to use a majority vote to decide on a property.

Still, counting peers over all crawls fails to estimate the value of a property for a \emph{typical} \ac{dht} graph.
Specifically, it overcounts nodes regenerating their peer ID and churning nodes.
It can be tackled by considering each crawl as a separate snapshot of the network, for which a property is derived, and then average over all crawls.
It is not sufficient to examine a single crawl, as this misses nodes that are offline during that point in time.
As such, we propose \emph{Average over Crawls, Unique Nodes} (A-N),
which assigns each peer a value per crawl and averages over all crawls,
which, for the example dataset, produces DE=$0.5$, US=1.
This is intuitively correct:
There is one stable node, probably situated in the US, and one node with 50\% uptime in Germany.

The single-crawl average number of nodes obtained through is A-N %
$25771.6$%
.
the number of unique peer IDs in aggregate over all %
$101$%
 crawls is %
$53898$%
.
A peer advertised an average of %
$1.82$%
 non-local IP addresses across all crawls,
which results in %
$86064$%
 unique non-local IP addresses (G-IP).

\paragraph{Bitswap logs}
Content retrieval in \ac{ipfs} always starts with a provider discovery phase.
This, currently, uses a local 1-hop broadcast via \bs{} to connected neighbours and subsequently searches the \ac{dht}.
Using the monitoring infrastructure described in~\cite{balduf2022monitoring},
we can monitor this discovery traffic of a large portion of the network via \emph{monitors}.
These are modified Go \ac{ipfs} implementations with unbounded connection capacity, which log all incoming \bs{} traffic to disk.
We run one \bs{} monitoring node which continuously collected data from August 2022 to May 2023.
We refer to this dataset as the raw, unmodified \emph{\bs{} traces}.
The traces are a subset of all \bs{} traffic in the network, because the monitor, albeit with unbounded connection capacity, is not connected to everyone in the network.
Additionally, as explained in \Cref{sec:background:content_retrieval}, only the initial provider discovery request is broadcast via \bs{}.
Lastly, we only see locally broadcast requests, not unicast responses.

In addition to the raw, unmodified \bs{} traces, we process the data further to obtain a daily sample of requested \acp{cid}.
For that, we aggregate all request traffic for a day and extract all requested \acp{cid}.
These are then deduplicated and a fixed amount of 200k is randomly sampled.
We refer to this dataset as \emph{daily sampled \bs{} \acp{cid}}.

\paragraph{Hydra-booster logs}
\begin{sloppypar}
We set up a modified version of Hydra-booster~\cite{hydra-booster} to collect the IPFS DHT traffic.
The Hydra-boosters acts as a DHT server with multiple, virtual peer IDs co-located on a single virtual machine.
We use Hydra-booster with 20 virtual peer IDs and modify it to write all the incoming DHT requests to disk.
We log the timestamp, the sender's peer ID and IP address, the type of the request, and the target key (peer ID or CID depending on the message type).
We also log the proxy DHT server if the original sender uses NAT traversal mechanisms (\Cref{sec:background}).
Apart from collecting the traffic, our software acts as a regular DHT server following the IPFS specification.
\end{sloppypar}

We collect traffic over two time periods:
from August 2022 to November 2022 and from February 2023 until April 2023 resulting in 290M messages.
Based on our observation, an average DHT query contacts 50 different nodes and the network contains 25,000 DHT servers. We thus estimate that our Hydra node captures around $4\%$ of the entire IPFS DHT traffic.
\paragraph{Provider Records} 
A peer can initiate a \dhtFindProviders{\cid} operation to retrieve provider records for CID \cid.
This operation collects up to 20 providers for a given \ac{cid} \cid,
terminating when either 20 providers of \cid have been found or all the resolvers (\ie 20 closest peers to \cid) have been queried for provider records of \cid.

We modified the \dhtFindProviders{\cid} implementation to terminate only when all the resolvers of \cid have been queried in order to retrieve all the provider records for the \acp{cid}.
Using this modified \ac{ipfs} \ac{dht} implementation, we retrieved all the provider records of the \acp{cid} from the daily sampled \bs{} \acp{cid} datasets starting from 23 April 2023 until 20 May 2023 for a total of 28 days.
The resulting dataset contained 5.6 million CIDs and their provider records.
We retrieved the provider records of each daily set of \acp{cid} on the same day they were collected.
\paragraph{Gateways}
Public gateways translate between \ac{http} and \ac{ipfs}.
While their \ac{http} endpoints are public, their overlay IDs are generally unknown.
We can identify these gateways on the overlay network through our \bs{} monitoring infrastructure in combination with specially crafted requests.

To identify a gateway, we generate a unique, random piece of data, and store it on our monitoring nodes.
We can be reasonably certain that we are the only provider of this data in the network.
We then request this data through the \ac{http}-side of a public gateway.
This will trigger the usual discovery and request mechanisms, which, eventually, result in a request via \bs{} to our monitoring node.
From this request, we can learn the overlay ID and address of the gateway.

Notably, many large gateway providers operate multiple \ac{ipfs} nodes to serve their traffic reverse proxied and served from a single \ac{http} endpoint.
While we can only identify one of these nodes per request, repeating these probes over time results in the discovery of multiple overlay IDs.
Eventually, we can be relatively certain to have identified all operational gateway nodes.

Of the 83 HTTP endpoints listed in the public gateway list, we find 22 gateways that functioned at least once, and 119 unique overlay IDs associated with these.
These numbers are in line with the ones reported via the public gateway checker tool~\cite{gateway_checker}.
\paragraph{Active and Passive DNS}
To estimate the number of domain names utilizing IPFS for content delivery through the DNSLink mechanism, we employ active and passive DNS data sources.

Our active scanning input list comprises domain names collected in April 2023 from the following sources:
\begin{enumerate*}
	\item centralized Zone Data Service~\cite{icann-czds} for both legacy and new generic TLDs (e.g., \texttt{.org} or \texttt{.xyz})
	\item publicly available zone files of three country-code TLDs: \texttt{.se}, \texttt{.nu}~\cite{se}, \texttt{.ch}~\cite{ch}
	\item Google Certificate Transparency logs~\cite{calidog}
	\item Tranco top popularity domain name ranking~\cite{tranco}, and
	\item passive DNS data kindly provided by SIE Europe~\cite{sieeurope}.
\end{enumerate*}

\begin{sloppypar}
We filter the domain names using Mozilla's Public Suffix list~\cite{publicsuffix} and retain only the root domain names (\eg, \texttt{example.com} or \texttt{example.com.uk}).
Next, using the \texttt{zdns}~\cite{zdns} scanner and Cloudflare Public DNS,
we send DNS \texttt{SOA} (Start of Authority) requests to determine registered domain names while excluding those resulting in an \texttt{NXDOMAIN} response code, indicating non-existing domains.
Our resulting list consists of 286M root domain names.
\end{sloppypar}

Administrators need to configure a DNS \texttt{TXT} record on a dedicated subdomain starting with the \texttt{\_dnslink} label to indicate the CID or the IPNS key's hash value associated with the domain.
Hence, for each domain name, we append the \texttt{\_dnslink} label (e.g., \texttt{\_dnslink.example.com} or \texttt{\_dnslink.example.com.uk}) and perform an active scan to retrieve the \texttt{TXT} records, verifying if they contain properly formatted DNSLink entries.
We actively scan for DNS \texttt{A} resource records on domains with valid DNSLink entries to ascertain whether the owner has configured a public IPFS gateway or another proxy server.
Note that our measurements do not include the identification of subdomains using IPFS for content delivery. The \texttt{\_dnslink} prefix can be added to any subdomain, such as \texttt{\_dnslink.subdomain.example.com}.

Finally, our objective is to compare the list of IP addresses of domain names using IPFS for content storage with the IP addresses of public gateways.
One approach is to perform active scans by querying DNS \texttt{A} resource records for the domain names of public gateways~\cite{gateway_checker}.
However, this approach has a limitation because DNS servers may provide different responses based on the geographic location of the querying client.
To address this limitation, we leverage one month of passive DNS data provided by SIE Europe from March 2023.
From this data, we extract all the observed IP addresses associated with the domain names of public gateways.
It is important to note that if a gateway operator utilizes Anycast DNS~\cite{rfc7094},
where the same IP address is advertised on multiple nodes, measurements from a single location would not impact the results.

\paragraph{Ethereum Name Service}
We examine ENS records pointing to IPFS CIDs. To collect our ENS dataset, we employ a similar methodology as used in~\cite{xia2022challenges}. Resolver contracts maintain information regarding name mappings. Therefore, we start by compiling an exhaustive set of 16 resolver smart contracts from prior work and Etherscan~\cite{etherscan}, and extract and traverse through the full history of event logs using the Etherscan API.

We filter for the \texttt{setContenthash()} function call, as defined in EIP-1577~\cite{eip1577}, which allows for a content hash to be set as the value of the record. From these, we specifically filter for records pointing to \texttt{ipfs\_ns} records, finding a total of $20.6$k records. We attempt to resolve the CID for each record to fetch providers of the content, finding $16.8$k provider records, out of which there were $9$k unique IPs.

\section{The Network}
\label{sec:network}

We examine the IPFS overlay through the \ac{dht} crawl dataset.
We discover an average of %
$25771.6$%
 peer IDs per crawl,
of which %
$17991.4$%
 are crawlable.
Over the entire dataset we observe %
$53898$%
 peer IDs
and %
$86064$%
 unique IP addresses.

\paragraph{Cloud Nodes}
First, we investigate the ratio of nodes hosted on major cloud providers in contrast to the number of non-cloud nodes.
Similar to a previous study by Trautwein \etal~\cite{trautwein2022design}, we employ the Udger IP database~\cite{udger}, which maps IP addresses to known cloud providers.
If there are no entries for a given address in the database, we mark it as non-cloud.
For peers announcing multiple cloud IP addresses, we assign the majority provider.
If a peer announces both cloud and non-cloud IP addresses, we assign it a \texttt{BOTH} label.

\begin{figure}[ht]
	\centering%
	\includegraphics[width=\linewidth]{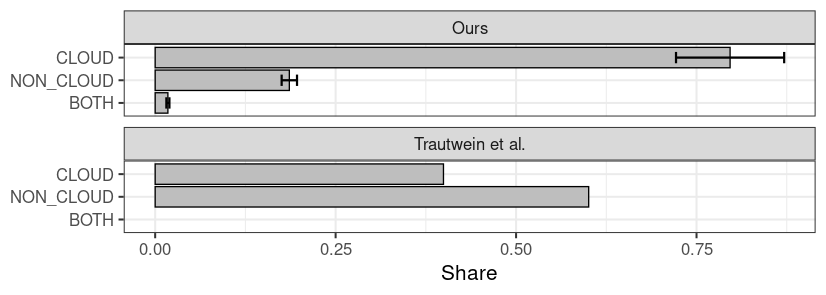}%
	\vspacehax{-0.3cm}%
	\caption[Participants of the IPFS DHT by Cloud Status, Comparison Between Counting Methodologies]{
		Participants of the IPFS DHT by Cloud Status, Comparison Between Counting Methodologies.
	}%
	\label{fig:topology_selected_counting_methodologies_cloud_status_overview}
	\vspacehax{-0.3cm}%
\end{figure}

\Cref{fig:topology_selected_counting_methodologies_cloud_status_overview} shows the ratio of cloud nodes found in the DHT.
Surprisingly, we discover a strong reliance on the cloud infrastructure.
An average of %
$20300$%
 (%
$79.6$\%%
) of the nodes are hosted in data centers,
with only %
$4737$%
 (%
$18.6$\%%
) non-cloud nodes.
Crawls of the \ac{ipfs} \ac{dht} only enumerate \ac{dht} servers, which require a public IP address.
The move towards cloud nodes can thus be explained by IPFS users having their regular machines hidden behind NAT.
However, such strong reliance on the cloud threatens the decentralization property of the system as the \ac{dht} forms the core of the platform.

Our results contrast with a previous study from 2022~\cite{trautwein2022design} showing less than 3\% of cloud nodes in the DHT.
We discovered that the inconsistency is due to a difference in aggregation and counting methodology (\cf \Cref{sec:methodology_datasets}) and a difference in the frequency, and thus number, of crawls.
The previous study creates a set of all unique IP addresses found across a large number of crawls (regardless of their relationship with the peer IDs) and then performs cloud provider attribution.
This results in %
$34375$%
 (%
$39.9$\%%
) addresses on cloud providers,
and %
$51689$%
 (%
$60.1$\%%
) non-cloud addresses.
We argue that this approach does not reflect the actual, typical state of the network.
As we show later, non-cloud IPFS nodes tend to be short-lived and frequently change their IP addresses,
artificially inflating their share in the network.

\begin{figure}[ht]
	\centering%
	\includegraphics[width=\linewidth]{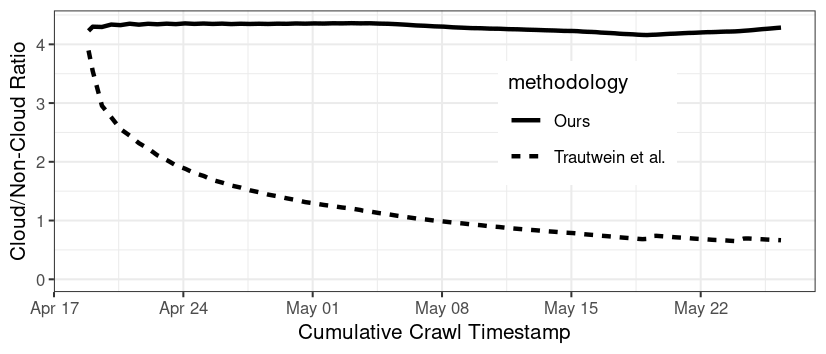}%
	\vspacehax{-0.3cm}%
	\caption[Ratio of Cloud to Non-Cloud Nodes as a Function Over Cumulative Successive Crawls]{
		The ratio of Cloud to Non-Cloud Nodes as a Function Over Cumulative Successive Crawls, comparison between counting methodologies.
	}%
	\label{fig:topology_cloud_non_cloud_ratio_cumulative_crawls}
	\vspacehax{-0.3cm}%
\end{figure}

We further showcase this phenomenon in \Cref{fig:topology_cloud_non_cloud_ratio_cumulative_crawls} showing the ratio of cloud to non-cloud nodes as a function of the number of aggregated crawls using both methodologies.
Using the approach from Trautwein \etal~\cite{trautwein2022design} makes the ratio of non-cloud nodes increase with the number of aggregated crawls.
This is because non-cloud nodes frequently rotating their IPs and churning nodes are counted multiple times.
On the other hand, aggregating the ratio of cloud to non-cloud nodes using our approach result in steady ratio values.
For the remainder of this section, we show results using both approaches to showcase how different views of the network can be obtained.


\begin{figure}[ht]
	\centering%
	\includegraphics[width=\linewidth]{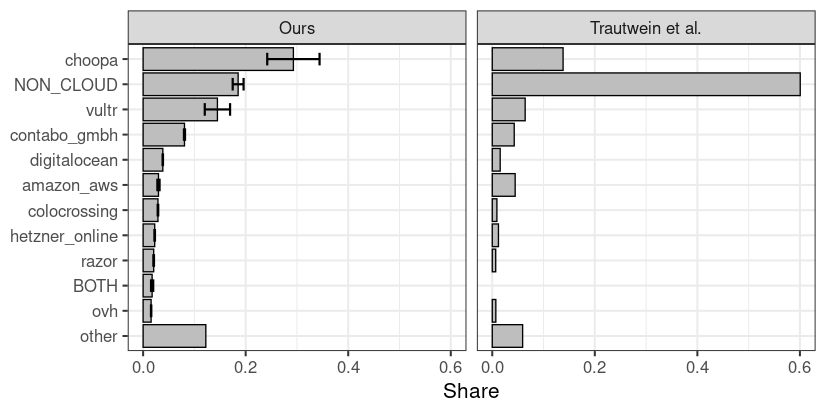}%
	\vspacehax{-0.3cm}%
	\caption[Nodes of the IPFS DHT Graph by Cloud Provider]{
		Nodes of the IPFS DHT Graph by Cloud Provider.
	}%
	\label{fig:topology_selected_counting_methodologies_cloud_status_detailed}
	\vspacehax{-0.3cm}%
\end{figure}

\paragraph{Cloud Providers}
In \Cref{fig:topology_selected_counting_methodologies_cloud_status_detailed}, we take a closer look at the cloud providers hosting IPFS nodes.
We find that the majority (%
$7492$%
, %
$29.3$\%%
) of nodes are hosted on \emph{choopa}
and the three main providers host %
$13259$%
 (%
$51.9$\%%
) of nodes.
This is a much stronger dependency on a single cloud provider than for other decentralized networks such as Mastodon,
where only 6\% of the nodes are hosted on Amazon AWS~\cite{raman2019challenges}.
Interestingly, using the alternative methodology~\cite{trautwein2022design} reduces \emph{choopa}'s share to %
$13.8$\%%
.
Furthermore, some providers such as \emph{digital ocean} have a relatively lower share using this methodology.
This might suggest that nodes hosted on these providers rotate their IP addresses less frequently.

\begin{figure}[ht]
	\centering%
	\includegraphics[width=\linewidth]{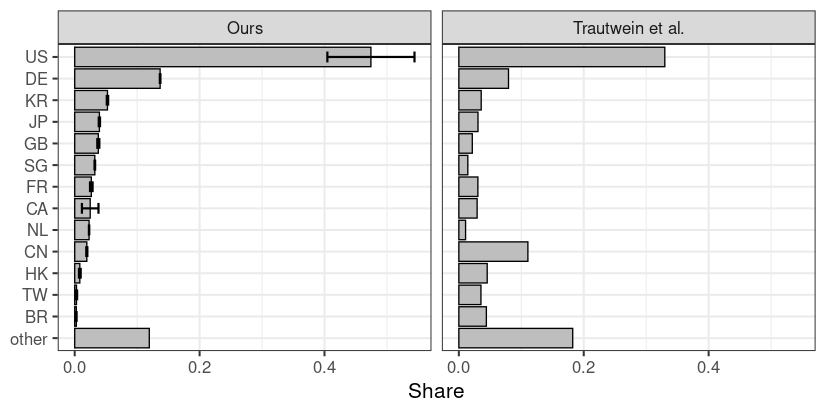}%
	\vspacehax{-0.3cm}%
	\caption[Nodes of the IPFS DHT Graph by Origin Country]{
		Nodes of the IPFS DHT Graph by Origin Country.
	}%
	\label{fig:topology_selected_counting_methodologies_geolocation}
	\vspacehax{-0.3cm}%
\end{figure}

\paragraph{Geolocation}
Next, we investigate the geospatial distribution of nodes in the overlay (\Cref{fig:topology_selected_counting_methodologies_geolocation}).
For this, we geolocate each node's addresses using the MaxMind GeoLite2~\cite{maxmind} database.
The majority of nodes are situated in the United States (%
$47.4$\%%
),
Germany (%
$13.7$\%%
),
and Korea (%
$5.2$\%%
)
while only %
$13.3$\%%
 are located outside the top 10 countries.
Note that this is independent of our vantage point, because we crawl the entire \ac{dht}.
This is in contrast to the results obtained via~\cite{trautwein2022design},
where the majority of peers reside in the United States (%
$33.0$\%%
),
China (%
$11.1$\%%
),
and Germany (%
$8.0$\%%
),
with non-top-ten countries accounting for %
$22.9$\%%
.
This is caused by short-lived IPs located in less represented countries that change frequently between crawls.

\begin{figure}[ht]
	\centering%
	\includegraphics[width=\linewidth]{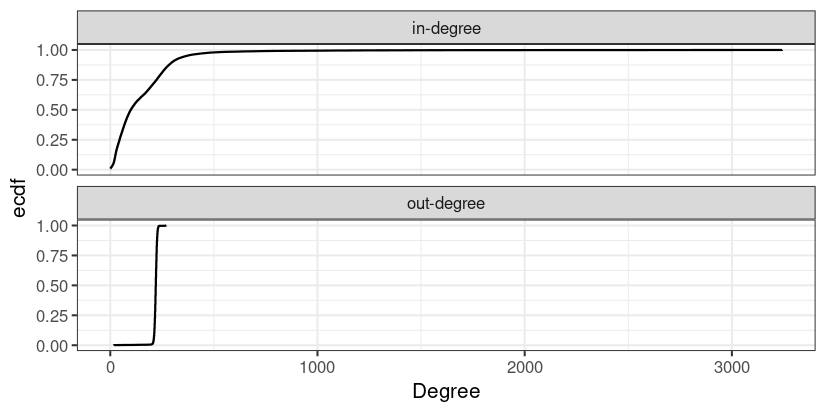}%
	\vspacehax{-0.3cm}%
	\caption[Degree Distribution of Nodes of the IPFS DHT Graph, Cumulative Density Function]{
		Degree Distribution of Nodes of the IPFS DHT Graph, Cumulative Density Function.
	}%
	\label{fig:topology_single_crawl_degree_distribution}
	\vspacehax{-0.3cm}%
\end{figure}

\paragraph{Node degree}
Using our \ac{dht} crawl dataset, we recreate the topology of the overlay network.
Out of %
$25771.6$%
 \ac{dht} servers,
$17991.4$%
 responded to our crawl requests, on average.
For those nodes, we learn their complete $k$-buckets, \ie, all outgoing \ac{dht} connections, which correspond to the edges in our graph.
The incoming connections are not directly available in the $k$-buckets, we thus estimate every node's in-degree by their presence in other peers' buckets.
\michal{TODO: need to discuss the in/out connections in the background section.}
This undercounts the true in-degree, as we cannot crawl all nodes.
The results of these investigations on a single graph from %
May \nth{12} 2023%
 are shown in \Cref{fig:topology_single_crawl_degree_distribution}.

The out-degree of nodes generally lies within a small band, which is ultimately dictated by the parameter $k$.
Even though \ac{ipfs} holds a large number of potentially unstructured connections,
only a subset of those, accounting for the structured Kademlia overlay, is stored in fixed-size buckets~\cite{henningsen2020mapping}.

The in-degree, on the other hand, is not dictated by Kademlia's $k$-buckets.
These are a subset of all connections of a node,
which are limited through a connection manager,
attempting to keep between 600 and 900 open connections.\footnote{These are preconfigured values, which were changed between \ac{ipfs} releases.}
Nodes may increase this value to improve the chances of discovering content providers through \bs{} (\cf \Cref{sec:background}).

We observe a few high in-degree nodes, pointing to those nodes having a high number of connections in general.
These nodes are contacted more often for \ac{dht} walks and perform central functionalities for the network.
This creates points of centralization, as outages within highly connected nodes will have a disproportionate impact on the overall graph structure~\cite{albert2000error}.
Out of the top 10 in-degree nodes, two are running a modified client by Filebase~\cite{filebase}, a pinning service, the others are seemingly regular go-ipfs v0.11 nodes, out of which 8 are hosted on Amazon AWS.
In general, though, most nodes have an in-degree of less than $\approx 200$, with the \nth{90} percentile being below $\approx 500$.

\michal{If we have data on what those nodes are, we can try to argue that a few platforms (e.g., NFT storage, handle the majority of the traffic.)}

\paragraph{Resistance to node removals}
Finally, we investigate the impact of node removals to analyze the tolerance to attacks and failures~\cite{yagan2013conjoining, joglekar2018online, raman2019challenges, henningsen2022empirical,albert2000error}.
We select a random crawl from %
May \nth{12} 2023%
with %
$24414$%
 total
and %
$16676$%
 crawlable nodes, which we interpret as an \emph{undirected} graph.
This allows \emph{all} observable connections of a node (\ac{dht} inbound and outbound) to be used for communication,
including \bs{}.

\begin{figure}[ht]
	\centering%
	\includegraphics[width=\linewidth]{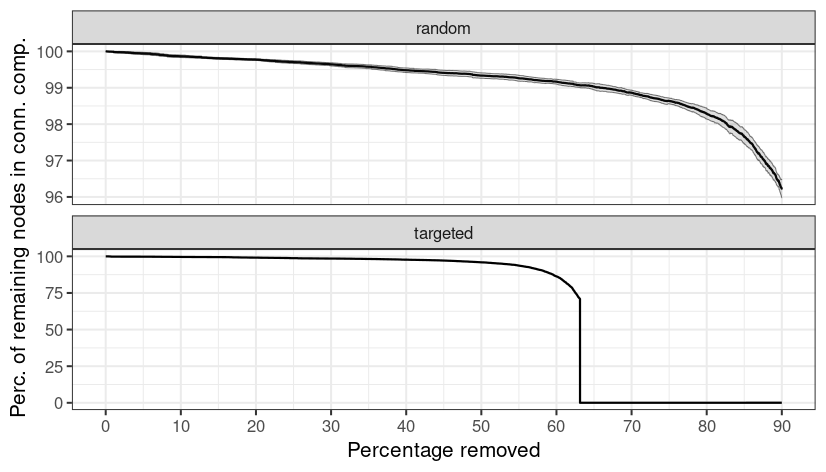}%
	\vspacehax{-0.3cm}%
	\caption[Resilience of the undirected IPFS DHT Graph to Random and Targeted Node Removals]{
		Resilience of the undirected IPFS DHT Graph to Random and Targeted Node Removals.
		\michal{Maybe the same y-axis for easier comparison?}
		Note the truncated vertical axis.
	}%
	\label{fig:topology_removals_ipfs_full_random_and_targeted}
	\vspacehax{-0.3cm}%
\end{figure}

We apply two different strategies of node removals to the graph:
\begin{enumerate*}
	\item Random, which picks a node at random, and
	\item targeted, which picks the node with the highest degree.
\end{enumerate*}
After each removal, we compute the connected components of the graph and count what portion of the remaining nodes is part of the largest~\cite{henningsen2022empirical}.
We repeat the random removal $10$ times to be able to show a $95\%$ confidence interval around the mean.
The results of this are shown in \Cref{fig:topology_removals_ipfs_full_random_and_targeted}.

The network is very robust to random removals.
The largest component spans $96\%$ of remaining nodes even after randomly removing $90\%$ of nodes.
This has been attributed to networks being scale-free~\cite{albert2000error}, \ie,
having a skewed distribution of node degrees.
This is commonly found in structured \ac{p2p} networks~\cite{salah2014characterizing}.
These results compare favourably to an earlier study done on \ac{ipfs} in 2019~\cite{henningsen2022empirical}, which finds that only $\approx 50$\% of nodes remain connected at this point.

Targeted removal is more effective in disconnecting nodes from the system.
The network ends up completely partitioned into components of size $1$ after $\approx 60\%$ of nodes were removed.
This, still, points towards good resistance to node removals:
A recent study on the resilience of the Mastodon social graph to targeted removal of users~\cite{raman2019challenges} found that removing even $\approx 10$\% of central user accounts leaves the majority of users outside the largest connected component.
For the Twitter graph, this occurs at $\approx 30$\%.
Our results compare favourably even to the earlier experiment on the \ac{ipfs} \ac{dht}, which finds complete partitioning after $\approx 40$\% of nodes were removed.

It thus seems that the \ac{ipfs} overlay, being very robust to random failures and only somewhat susceptible to targeted attacks,
possesses properties both of structured~\cite{salah2014characterizing} and unstructured~\cite{stutzbach2005evaluating} networks.
This is in line with prior considerations on the robustness of the network~\cite{henningsen2020mapping,balduf2022monitoring}.
The network has even improved in robustness since the study performed in 2019~\cite{henningsen2022empirical},
which found that unconnectable leaf nodes can be removed easily with targeted attacks.
These leaf nodes have been largely eliminated from the \ac{dht} in recent versions due to the differentiation between \ac{dht} servers and clients.
Note, however, that we simplified the graph to be undirected, which allows \bs{} to be used on all edges,
but ignores that \bs{} broadcasts only travel one hop.
As such, even though the network stays mostly connected, it is not guaranteed that content stays available equally.
A more nuanced analysis is left for future work.

\section{The Traffic}\label{sec:traffic}
In this section, we investigate the traffic generated by the IPFS network. We use and compare our \bs and Hydra-booster datasets. We classify the DHT traffic into content-related \textit{downloads} (\eg requesting providers for a specific CID) and \textit{advertisements} (\eg announcing a new provider for a specific CID). We ignore all the other types of messages (\eg nodes joining the network). We observe 290M messages where download-related traffic represents 57\%, advertisement-related traffic 40\% and other types of messages 3\%. Importantly, our traffic traces include the NAT-ed peers (\ie DHT clients) that were not visible in the previous section. Although Hydra-boosters receive traffic from DHT clients, their traffic is not distinguishable from the traffic by DHT servers. In \Cref{sec:provRecords}, we present a better picture of how NAT-ed DHT clients impact content hosting in \ac{ipfs}. 

\begin{figure*}[ht]
	\centering
	\minipage{0.31\textwidth}
	\includegraphics[width=\linewidth]{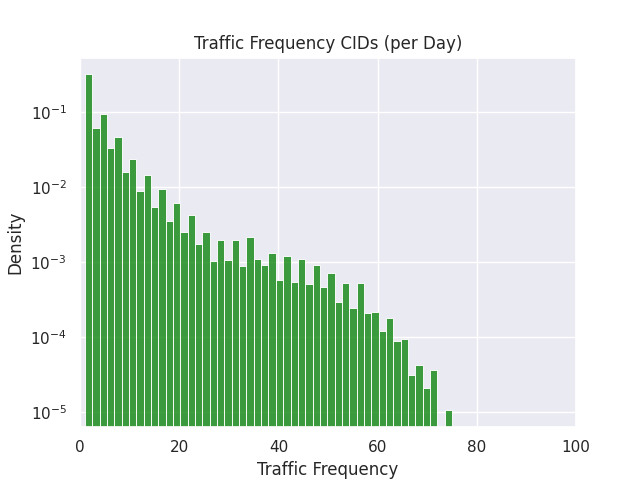}
	\caption*{(a) CID}
	\endminipage\hfill
	\minipage{0.31\textwidth}
	\includegraphics[width=\linewidth]{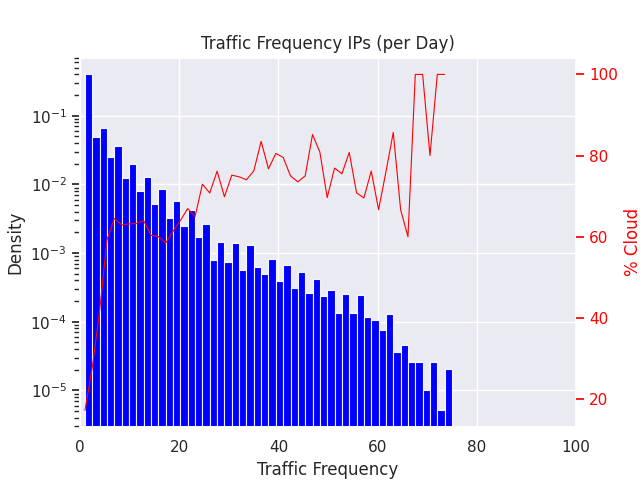}
	\caption*{(b) IP}
	\endminipage\hfill
	\minipage{0.31\textwidth}%
	\includegraphics[width=\linewidth]{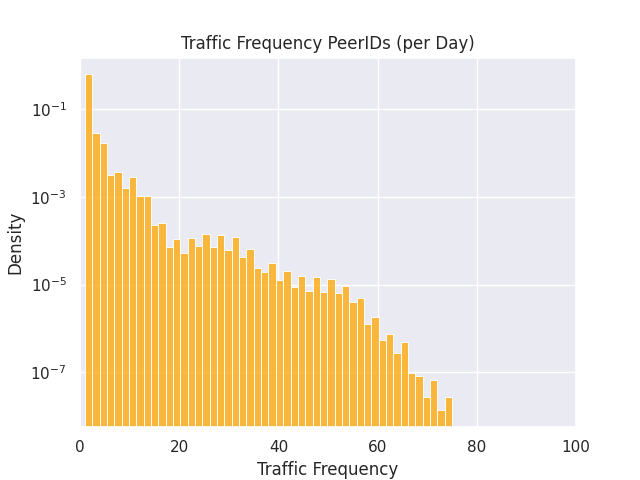}
	\caption*{(c) PeerID}
	\endminipage
	\vspacehax{-0.3cm}%
	\caption{Request frequency per identifier (in days seen). Note the y-axis log scale.}
	\label{fig:hydra_temporal}
	\vspacehax{-0.3cm}%
\end{figure*}

\paragraph{General metrics}
First, we explore the temporal properties of the IPs, peerIDs and CIDs found in our Hydra-booster logs (\Cref{fig:hydra_temporal}).\footnote{We exclude \bs logs from this analysis due to the sheer volume of data. Analyzing a random sample of the logs would not preserve the temporal properties.}
We define frequency as the number of different days where we observe the item.
The vast majority of the CIDs are downloaded or advertised only for 1-3 days. This suggests that IPFS is mostly used for direct content transfer rather than persistent storage.

The majority of IPs and peerIDs are also short-lived. The result for the IPs suggests IP rotation and further explains result differences for two methodologies in \Cref{sec:network}. We observe a large portion of cloud nodes in general but their share increases for IPs seen over many days. The short-lived nature of the peerIDs is also surprising. By default, IPFS clients keep their peerIDs across multiple restarts. This suggests that many users use the network for a single interaction.

\begin{figure}[ht]
	\centering
	\includegraphics[width=\linewidth]{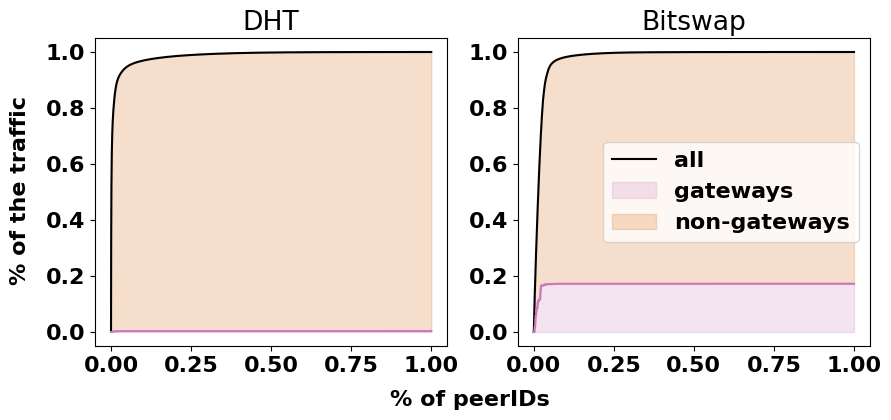}
	\vspacehax{-0.3cm}%
	\caption{DHT/Bitswap peerID simplified Pareto chart.}
	\label{fig:id_cdf}
	\vspacehax{-0.3cm}%
\end{figure}

\paragraph{ID centralization}
We then investigate the centralization of the traffic through the lens of the peerIDs (\Cref{fig:id_cdf}) and distinguish between gateway and non-gateway nodes based on our \emph{gateway dataset}. For both \bs and DHT, we observe high centralization of the traffic far beyond the Pareto $20\%$-$80\%$ principle. $5\%$ of the most active peerIDs are responsible for almost $97\%$ of the traffic.
The gateway ratio differs significantly for both protocols ($\approx 1\%$ for DHT, $\approx 18\%$ for bitswap). We suspect that this is caused by a large number of \bs connections maintained by each gateway and fixed links to the industrial content providers such as \emph{pinata} or \emph{nft.storage}. The gateways thus satisfy a vast majority of content requests via \bs and do not rely on the DHT for content resolution.

\begin{figure}[ht]
	\centering
	\includegraphics[width=\linewidth]{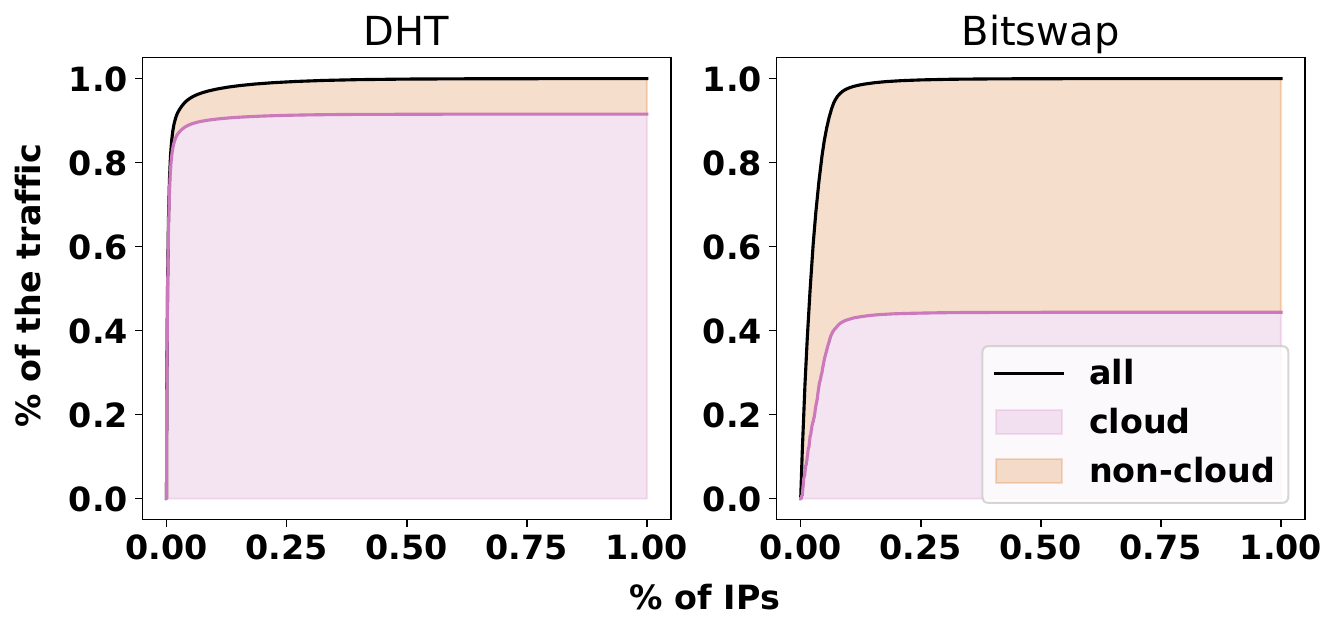}
	\vspacehax{-0.4cm}%
	\caption{DHT/Bitswap IP simplified Pareto chart.}
	\label{fig:ip_cdf}
	\vspacehax{-0.3cm}%
\end{figure}

\paragraph{IP centralization}
We repeat the centralization experiments this time looking at IPs and distinguishing between cloud and non-cloud nodes (\Cref{fig:ip_cdf}). We observe a similar, high centralization of the traffic with 5\% of the most active IP addresses responsible for almost 94\% of messages. For the DHT, cloud nodes are the most active ones in the network generating a staggering $\approx 85\%$ of the traffic. The non-cloud nodes, while similar in number, are much less active and are responsible for only $\approx 15\%$ of the traffic. The distribution is similar for both download- and advertisement-related traffic (omitted on the graph). The \bs logs show a much smaller but still significant share of cloud nodes ($\approx 42\%$). We explain this phenomenon in the paragraphs below.

\begin{figure}[ht]
	\centering
	\includegraphics[width=\linewidth]{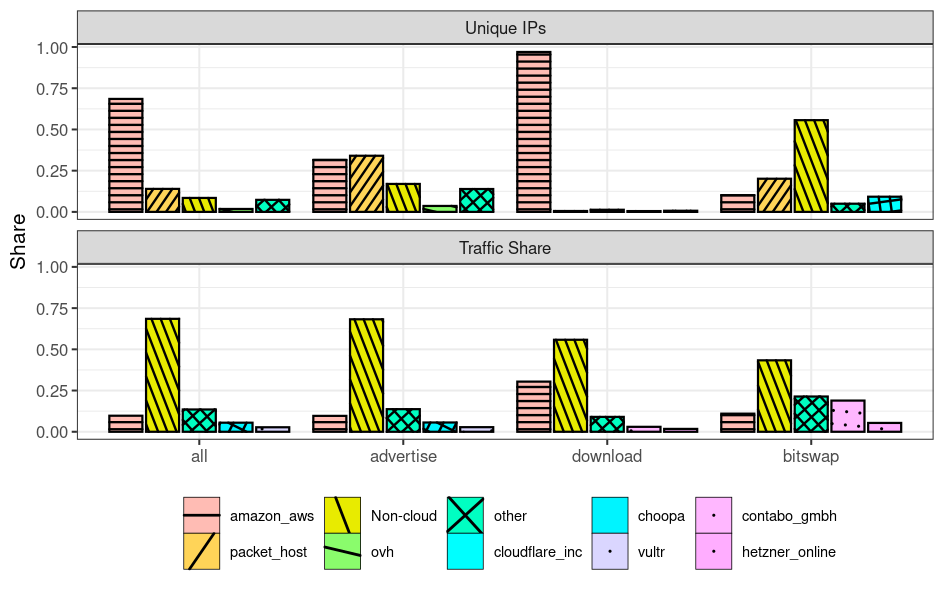}
	\vspacehax{-0.3cm}%
	\caption{Cloud per traffic type
	}%
	\label{fig:cloud_per_traffic_type}
    \vspacehax{-0.3cm}%
\end{figure}

\paragraph{Cloud providers} In \Cref{fig:cloud_per_traffic_type} (top graph), we analyze all IPs present in our logs distinguishing between traffic related to content downloading and advertising and indicating the most popular cloud providers. $35\%$ overall ratio of cloud-based nodes. This is a smaller ratio than found during the network crawls ($79\%$ in \Cref{sec:network}). This is understandable, as the network crawls do not include the nodes behind NAT. We observe a similar division across cloud providers with \emph{choopa, vultr} and \emph{contabo} being the most popular ones, but an increased share of nodes hosted at Amazon AWS. Surprisingly, the cloud-based nodes are more present in the traffic related to downloading files ($\approx 45\%$) than in the advertising traffic ($\approx 34\%$).

We repeat the analysis but this time we take into account the traffic generated by each IP address (\Cref{fig:cloud_per_traffic_type} bottom graph). Again, we observe that the cloud-based nodes are much more active and are responsible for $\approx 93\%$ of the traffic. The ratio goes up to $\approx 98\%$ for download-related traffic. The share of Amazon AWS raises to a staggering 68\% followed by packet host being jointly responsible for 82\% of the traffic.

\begin{figure}[ht]
	\centering%
	\includegraphics[width=\linewidth]{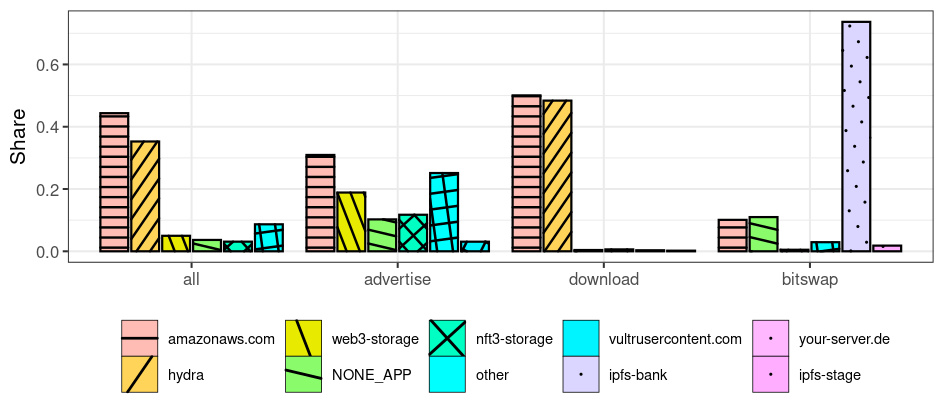}%
	\vspacehax{-0.3cm}%
	\caption{Platforms generating traffic based on reverse DNS lookups.
	}%
	\label{fig:traffic_apps}
	\vspacehax{-0.3cm}%
\end{figure}

\paragraph{IPFS-based platforms}
To complete the picture of the IPFS traffic, we analyze the applications/platforms responsible for the traffic. First, we obtain a set of peerIDs of Hydra-booster nodes operated by Protocol Labs and hosted on Amazon AWS. Those nodes were deployed in the network to speed up the DHT lookups. For all the non-Hydra IPs, we perform reverse DNS lookups (\Cref{fig:traffic_apps}).

The Hydra-booster nodes are responsible for $35\%$ of all the DHT traffic and $50\%$ of the download traffic. However, it is not visible in the content advertising traffic. When a Hydra-booster receives a DHT content resolution request, it acts as a regular DHT node. First, it checks its cache for the relevant provider records. If those are unavailable, the Hydra-booster returns a list of peers closer to the target CID. However, the Hydra-booster also initiates its own lookups for all the requested CIDs that are not found in the cache, trying to proactively fill the cache for future requests. This behaviour exposes an easy Denial of Service vector. Asking a Hydra-booster for non-existing content generates significant amounts of traffic in the network. We are unable to confirm whether such intentional attacks happened during our measurement period or whether Hydra-booster nodes simply amplify regular requests for, potentially non-existing, content. This explains the higher share of cloud-based nodes in the DHT compared to the \bs traffic in \Cref{fig:ip_cdf} and a higher share of Amazon AWS download-related traffic (\Cref{fig:cloud_per_traffic_type}).

On the other hand, a few storage platforms using IPFS (\emph{web3-storage} and \emph{nft3-storage}) dominate the DHT advertise-related traffic. Those platforms offer practical persistent storage over IPFS using cloud infrastructure and thus periodically advertise all their CIDs in the network and explain the high cloud usage in the advertise-related DHT traffic (\Cref{fig:ip_cdf}, \Cref{fig:cloud_per_traffic_type}). The \bs traffic is dominated by \emph{ipfs-bank} which is an HTTP gateway platform. Unfortunately, we were not able to discover the purpose of the remaining traffic originating from Amazon AWS. While convenient for the end users and increasing accessibility, those platforms significantly contribute to the centralization of the network.

\section{The content providers}\label{sec:provRecords}

\begin{sloppypar}
In this section, we focus on the providers of content on IPFS.
A provider record is a mapping of \ac{cid} to \emph{multiaddresses} --- a self-describing address format, \eg, /ip4/1.10.20.30/tcp/29087/ipfs/<peer ID>, that embeds provider's connectivity information and peer ID. We collect all the provider records of 5.6 million \acp{cid} over 28 days. As provider records may be stale, \ie a provider may have gone offline, we verified the reachability of the providers at the time of retrieving the provider record and ignored the unreachable ones.
\end{sloppypar}

\begin{figure}[ht]
	\centering
	\includegraphics[width=\linewidth]{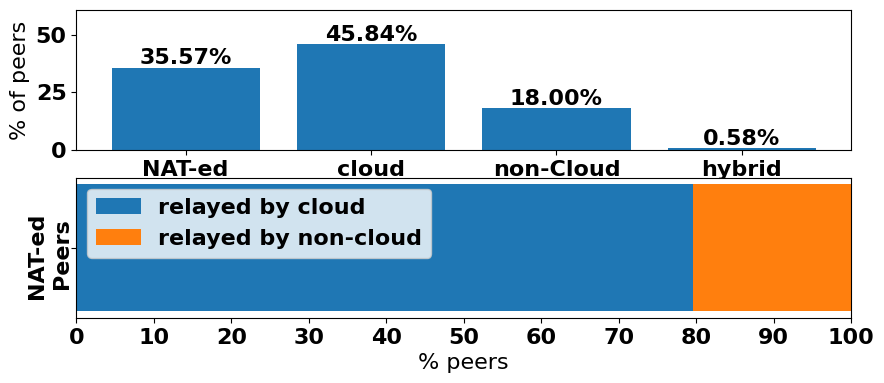}
	\vspacehax{-0.5cm}%
	\caption{Classification of providers.}
	\label{fig:prlogs_peer_clasification}
    \vspacehax{-0.3cm}%
\end{figure}
In IPFS, NAT-ed peers (\ie DHT clients) can still provide content. This is done through a circuit relay protocol~\cite{circuitRelay} where the NAT-ed peers keep a connection with a \emph{relay} (\ie a DHT server with a public IP) that reverse proxies the connection requests to make the NAT-ed peers reachable. When a NAT-ed peer advertises content, it provides a keyword \emph{circuit} and the relay's IP address in its multiaddress.%

\paragraph{Peer analysis} We categorize each unique provider peer ID based on its IP address(es) as one of: NAT-ed, cloud-based, non-cloud-based, and hybrid (\Cref{fig:prlogs_peer_clasification}). The NAT-ed peers account for a significant portion (35.57\%) of the providers. The cloud-based peers are the majority of the providers (45\%), while 18\% of the peers are non-cloud peers with public IPs. On the other hand, a very small percentage of the providers (\ie 0.58\%) had a mix of cloud and non-cloud IP addresses (\emph{hybrid} in~\Cref{fig:prlogs_peer_clasification}). Those peers either have two instances in both cloud and non-cloud nodes or have moved during the provider record collection as we take a snapshot of a content's providers only once. 

\begin{figure}[ht]
	\centering
	\includegraphics[width=\linewidth]{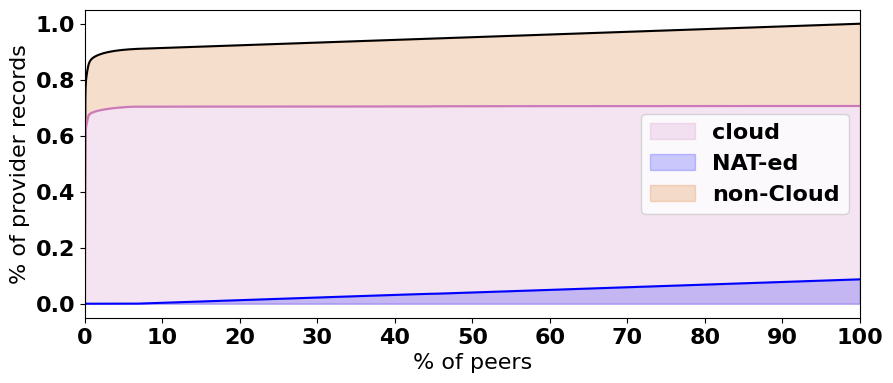}
	\vspacehax{-0.6cm}%
	\caption{A simplified Pareto chart of peer IDs.}
	\label{fig:prlogs_ip_providers}
    \vspacehax{-0.3cm}%
\end{figure}
In the bottom plot of \Cref{fig:prlogs_peer_clasification}, we present the distribution of relay nodes used by the NAT-ed peers. We observe that around 80\% of the NAT-ed peers use a cloud-based peer as a relay node. This means that a large portion of NAT-ed peers makes use of cloud-based nodes to provide content. 

\paragraph{Provider popularity} We look into the popularity of each provider in terms of the number of times each appears in the collected provider records (\Cref{fig:prlogs_ip_providers}). A small percentage of peers appear in a large percentage of provider records. Around 1\% of the peers appear as \textit{one of the} providers in approximately 90\% of the records. A large portion (\ie 70\%) of these popular providers are cloud-based, while NAT-ed peers appear in less than 8\% of the records. On the other hand, the non-cloud providers (with public IP addresses) appear in around 22\% of the provider records.

\begin{figure}[ht]
	\centering
	\includegraphics[width=\linewidth]{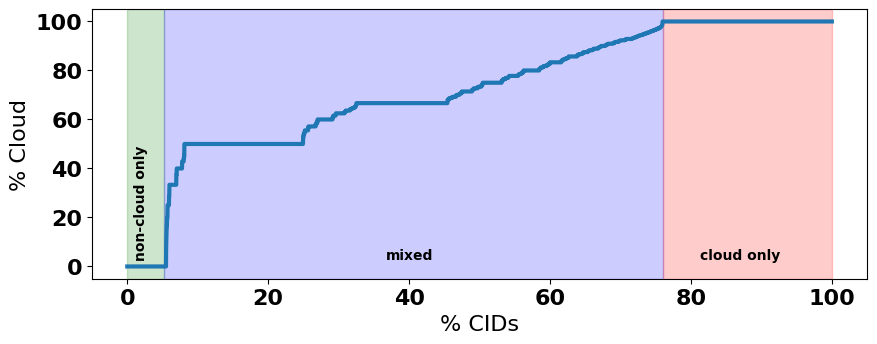}
	\vspacehax{-0.8cm}%
	\caption{CIDs classified based on their providers.}
	\label{fig:prlogs_cid_classification}
    \vspacehax{-0.3cm}%
\end{figure}

\paragraph{Content-level analysis} A single \ac{cid} can be hosted by multiple providers. We explore the properties of the content in terms of its reliance on cloud-based peers. For each content, we calculate the percentage of its cloud-based providers among all of its providers (shown as ``\% Cloud'' in \Cref{fig:prlogs_cid_classification}). In this calculation, we classified NAT-ed providers as non-cloud peers. Nearly 95\% of the content is provided by at least one cloud-based provider. At the same time, for 91\% of the content, at least half of the providers are cloud-based and 23\% of the content is provided only by cloud-based peers. On the other hand, an alternate interpretation posits that around 77\% of the CIDs have at least one non-cloud provider.

To sum it up, content hosting in IPFS is heavily reliant on cloud-based infrastructure. Cloud nodes are not only dominant in content hosting, but they also serve as relays for a large portion of NAT-ed content providers.

%

%

\section{The entry points}\label{sec:entry-points}
In this section, we investigate entry points to IPFS. Namely, HTTP Gateways and two systems mapping human-readable domain names to IPFS CIDs.\footnote{IPFS also provides IPNS - one more way of mapping human-readable names to CIDs. We skip this mechanism as it is internal for IPFS and is equivalent to regular CID fetching already covered in \Cref{sec:traffic}.}

\begin{figure}[ht]
\centering

\begin{subfigure}{\linewidth}
  \centering
  \includegraphics[width=\linewidth]{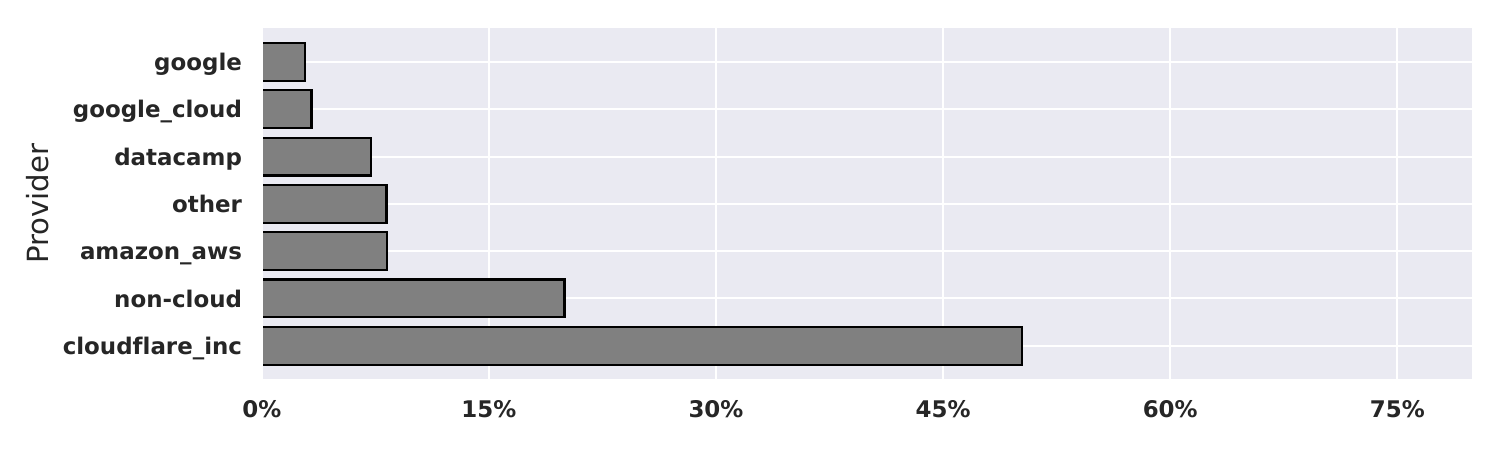}
  \caption{Cloud providers}
  \label{fig:dnslink:1}
\end{subfigure}

\begin{subfigure}{\linewidth}
  \centering
  \includegraphics[width=\linewidth]{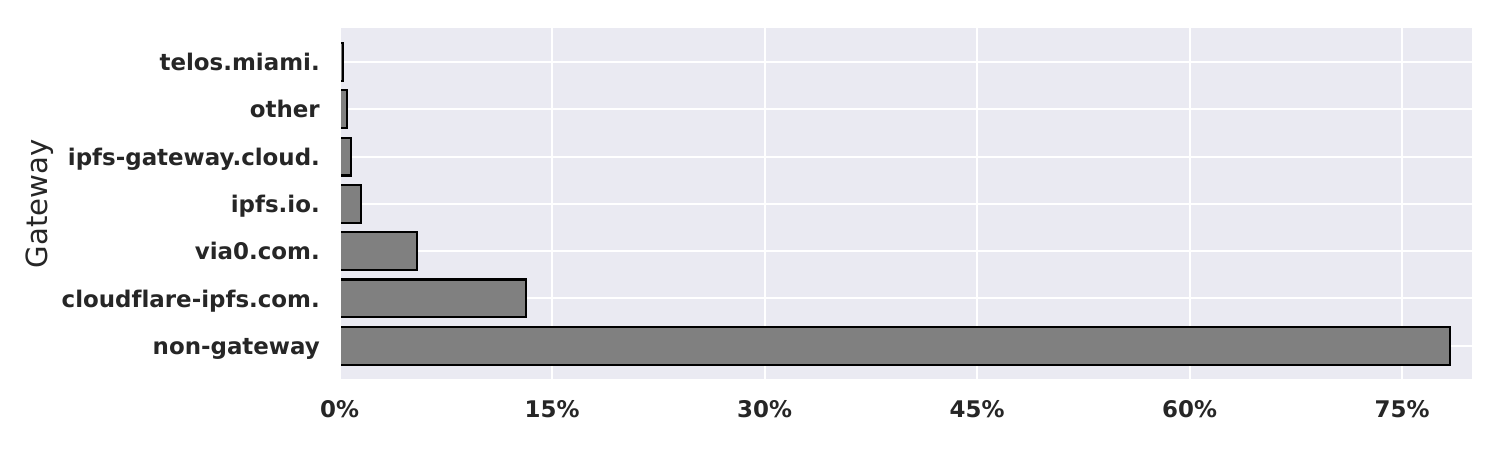}
  \caption{Gateways}
  \label{fig:dnslink:2}
\end{subfigure}

\caption{DNSLink statistics for records pointing to IPFS content providers}
\label{fig:dnslink}
\end{figure}

\paragraph{DNSLink}
DNSLink allows resolving domain names to CIDs using the DNS. However, each correct record must contain an HTTP gateway through which the CID can be fetched (\Cref{sec:background}). \Cref{fig:dnslink} presents the distribution of these IP addresses across cloud providers. Similarly to our previous measurements, only 20\% of the gateways are non-cloud nodes. This is understandable, as gateways require a public IP and high availability to operate efficiently. However, we observe different popularity of specific cloud providers. 50\% of the IP addresses are hosted by Cloudflare alone. Cloudflare actively supports IPFS, for instance, by operating one of the most popular public HTTP gateways that can be easily used for DNSLink.
Surprisingly, we observe only $~21\%$ of the IPs belonging to public gateway domains~\cite{gateway_checker}.\michal{Not sure how to comment on that. It might suggest that our dataset is not great.}

\paragraph{Gateways}
We investigate IPFS gateways themselves using the \emph{gateway ID} and \emph{passive DNS datasets}. We collect the domain names of public gateways and those referenced by DNSLink records. We distinguish between HTTP-facing (\ie accepting HTTP requests) and overlay-facing (\ie issuing requests to the IPFS network) IP addresses.

\begin{figure}[h]
	\centering%
	\includegraphics[width=\linewidth]{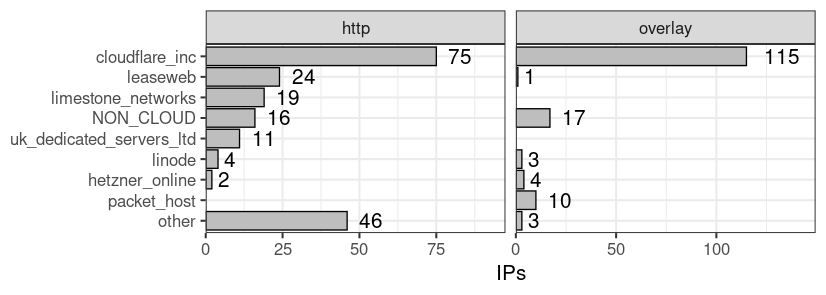}%
	\vspacehax{-0.3cm}%
	\caption[Unique IP Addresses of Gateway Frontends and Overlay Nodes by Cloud Provider]{
		Unique IP Addresses of Gateway Frontends and Overlay Nodes by Cloud Provider.
	}%
	\label{fig:gateway_discovery_cloud_providers}
    \vspacehax{-0.2cm}%
\end{figure}

\begin{figure}[h]
	\centering%
	\includegraphics[width=\linewidth]{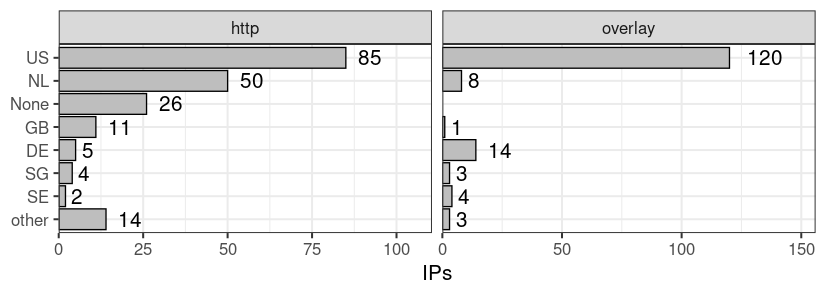}%
	\vspacehax{-0.5cm}%
	\caption[Unique IP Addresses of Gateway Frontends and Overlay Nodes by Geolocation]{
		Unique IP Addresses of Gateway Frontends and Overlay Nodes by Geolocation.
	}%
	\label{fig:gateway_discovery_geolocation}
    \vspacehax{-0.3cm}%
\end{figure}

Similar to the DNSLink data, we find heavy reliance on Cloudflare (\Cref{fig:gateway_discovery_cloud_providers}).
This is unsurprising for the \ac{http} frontends, as Cloudflare is commonly employed as a reverse proxy and protection service,
a practice equally often criticized for creating central points of failure~\cite{cloudflareoutage}.
Conversely, some of these gateways may actually be located in other autonomous systems, or hosted on different providers.
However, we also find a large number of IPs provided by Cloudflare on the overlay side.
It appears that these IPs are internal to Cloudflare, utilized to reverse proxy the \emph{overlay} connections of their Gateways.\footnote{We are currently confirming this with Cloudflare.}
A notable number of gateways are running on non-cloud systems.
This is commendable and potentially due to the open nature of the gateway ecosystem:
In principle, anyone can operate a gateway and add themselves to the public register~\cite{gateway_checker}.
Note that, with the exception of Cloudflare, gateway providers generally do not operate their own AS.

Geolocating both the HTTP frontend IP addresses as well as the overlay addresses using the MaxMind GeoLite2~\cite{maxmind} database,
we find that the majority of nodes are situated in the United States and Germany (\Cref{fig:gateway_discovery_geolocation}),
similar to the overall trends observed in the \ac{dht} (\cf \Cref{sec:network}).
The large number of frontend IP addresses situated in the Netherlands could be a result of our vantage point in Germany.
The geolocation of overlay addresses is unaffected by our vantage point.

\begin{figure}[ht]
\centering

\begin{subfigure}{\linewidth}
  \centering
  \includegraphics[width=\linewidth]{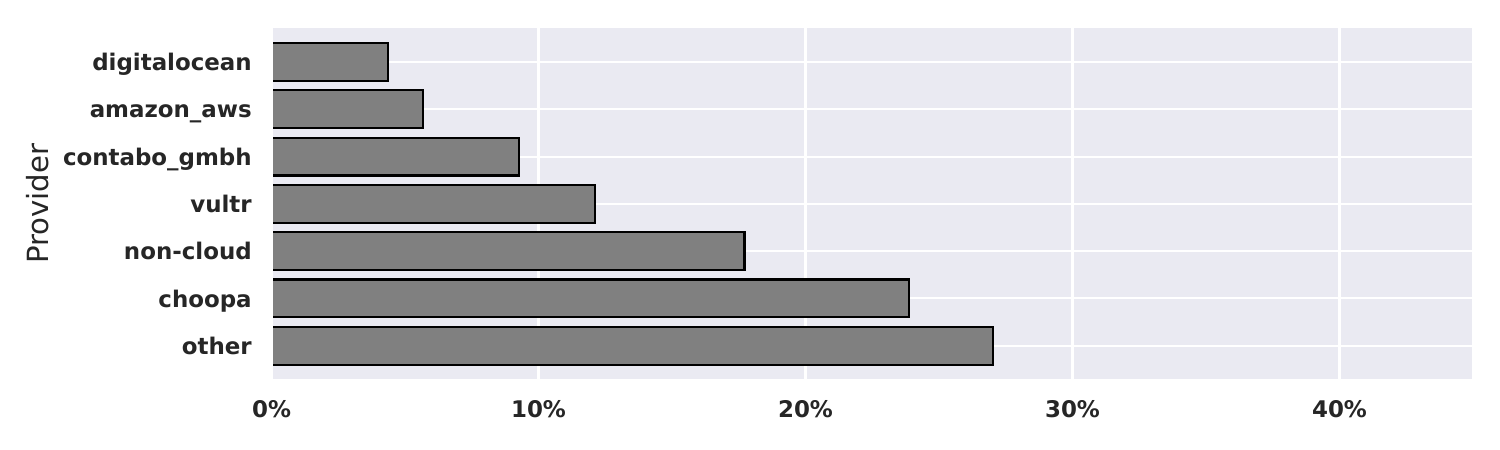}
  \caption{Cloud providers}
  \label{fig:ens:1}
\end{subfigure}

\begin{subfigure}{\linewidth}
  \centering
  \includegraphics[width=\linewidth]{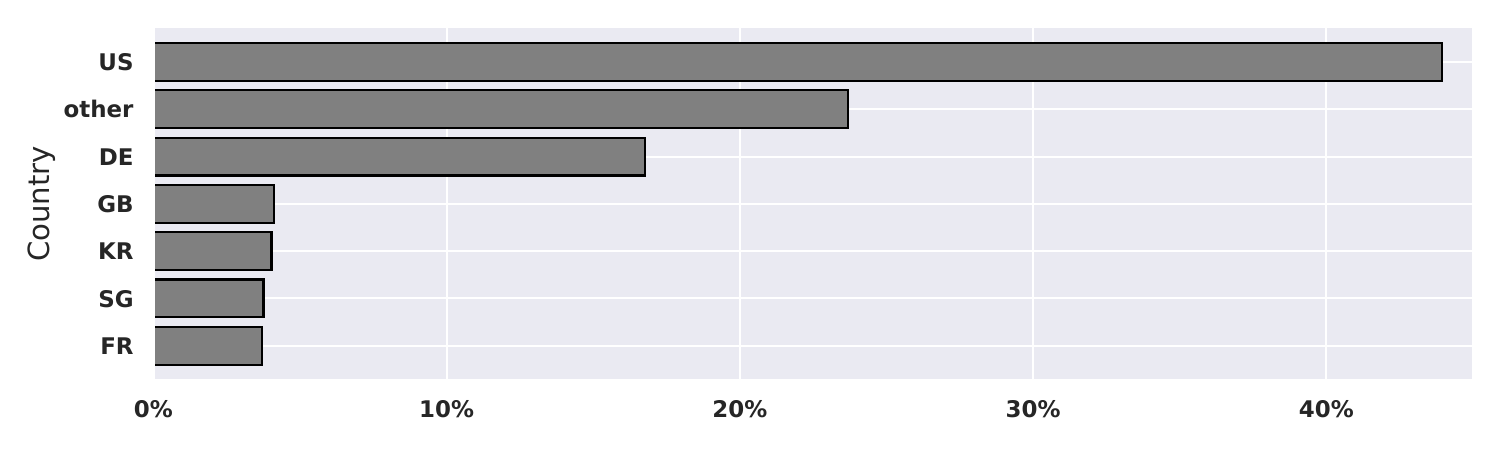}
  \caption{Geolocation}
  \label{fig:ens:2}
\end{subfigure}

\caption{Content provider statistics for IPFS content on ENS records (taking unique IPs)}\label{fig:ens}
\end{figure}

\paragraph{Ethereum Name Service}
Combining IPFS with ENS is a popular approach to scale content delivery in Web3. ENS records point to CIDs that can be later downloaded from IPFS. We extract a list of IPFS providers for each CID referenced by ENS and analyze them in the context of cloud providers (\Cref{fig:ens}.a). The results are consistent with the previous measurements: 82\% of the content is hosted on cloud nodes with the main providers being \emph{choopa}, \emph{vultr} and \emph{contabo}. 

We then investigate the geolocation of the providers (\Cref{fig:ens}.b) and observe that the majority of content ($60\%$) is concentrated in the US and Germany alone. Even with the ENS records being held on a blockchain, the actual retrieval of the referenced content is centralized and heavily dependent on a few cloud providers.

\section{Related Work}\label{sec:related}
\paragraph{DHT Measurements} Numerous studies investigated the state and performance of various DHTs implementation such as KAD \cite{salah2014characterizing,steiner2007global,steiner2007actively} or BitTorrent DHT~\cite{falkner2007profiling,crosby2007analysis,wolchok2010crawling,kaune2010unraveling}. In contrast, we focus on the multimodal analysis of multiple protocols used in IPFS.

\paragraph{IPFS}
We add to a growing body of research on IPFS~\cite{benet2014ipfs}. Henningsen \etal~\cite{henningsen2020mapping,henningsen2022empirical} develop a crawler for the \ac{ipfs} network we use for this work. Their DHT analysis from December 2019 finds lower overall robustness to node removal than our study.
We attribute this to improvements in the \ac{dht} since 2019, most notably the removal of unconnectable leaf nodes.
Balduf \etal~\cite{balduf2022monitoring} investigate privacy issues relating to unstructured \bs{} broadcasts. \cite{daniel2022passively} focuses on network participants and their churn. 
Daniel \etal~\cite{daniel2022ipfs} provide a comprehensive overview of the IPFS ecosystem and its components but without performing any measurements.
Trautwein \etal~\cite{trautwein2022design} gives an overview of the functionality of \ac{ipfs} and measures its client population over an extended period of time. 
Other studies describe security vulnerabilities leading to eclipse~\cite{prunster2022total} or censorship~\cite{genoneclipse} attacks or investigate how IPFS can be used to spread malware~\cite{patsakis2019hydras}.

\paragraph{Assessing Decentralization}
Networks, both in nature and communications, have been analyzed for their properties and robustness before.
In their seminal work, Albert \textit{et al.} showcased how many complex networks,
including the internet and social networks, are resistant to random failures due to being scale-free~\cite{albert2000error}.
More recently, such analyses have been applied to Diaspora~\cite{bielenberg2012growth}, and Mastodon~\cite{raman2019challenges} as well as the Bitcoin and Ethereum blockchains~\cite{gervais2014bitcoin,gencer2018decentralization}.
Sadly, and similar to our study, those papers usually report a higher level of centralization than expected for their respective platforms.
To the best of our knowledge, we are the first to analyze and assess the decentralization of IPFS in a comprehensive way.

\section{Discussion and conclusion}\label{sec:conclusion}

We find evidence of a heavy IPFS reliance on cloud infrastructure that is visible in the network topology, generated traffic, content provider records and its entry points. This dependency threatens the core design goals of the IPFS such as censorship resistance, robustness and openness.

One of the main challenges of IPFS remains the inability of NAT-protected nodes to fully participate in the network.
It limits the number of full DHT participants, %
putting more pressure on the public-IP nodes.
Furthermore, hosting content using proxies increases centralization and reliance on cloud nodes.
IPFS supports a NAT-traversal protocol that requires assistance from a public-IP node only during the connection setup~\cite{seemann2022decentralized,dcutr}, but NAT-punching clients still function as DHT clients only.
However, in the long run, the wider deployment of IPv6, and thus the removal of IPv4 NAT, seems like a more sustainable solution.

The Hydra-booster nodes deployed by Protocol Labs were supposed to speed up the content resolution by acting as a large provider records cache and DHT query speed amplifier. They are thus cloud-based and operated by a single entity. On the other hand, the presence of Hydra-boosters is not an entirely bad thing as the DHT can still function properly, should the Hydra-booster nodes disappear. 

A more concerning idea is the recent introduction of network indexers that are entirely hosted in the cloud~\cite{indexers}. The indexers gather information about all the content stored on IPFS and can resolve it much faster than the current DHT lookups. Content resolution is a core functionality of the platform and its control by a single entity gives its operator the power to block content (\eg when pressured by the government). 

In general, cloud-based resolution is always faster than decentralised lookup. Its deployment may be thus important for multiple latency-sensitive applications. At the same time, we strongly advise keeping the DHT as a fallback resolution mechanism to maintain the decentralization of the network. More research on more efficient DHTs (or similar resolution networks) is also needed to close the performance gap between the two solution classes.

Protocol Labs introduced multiple commendable solutions (\eg Brave integration, HTTP Gateways, DNSLink) making the network easy to use for non-tech-savvy users and boosting its adoption. However, if done incorrectly, they can also introduce centralization. For instance, Brave users can currently choose between a self-hosted IPFS node and a default, cloud-based gateway. Changing the default gateway to a random one supported by a dynamic, permissionless discovery system could maintain simplicity while avoiding reliance on cloud infrastructure.

Currently, the IPFS network is used as a file transfer system rather than decentralized storage, based on the short lifetime of data items.
At the same time, storage persistence is provided by a handful of centralized, cloud-based providers. Mechanisms such as Filecoin~\cite{psaras2020interplanetary} could help to solve this problem by providing incentives for storing content. However, it is unclear whether decentralized storage nodes can compete with the service quality and prices of dedicated cloud providers. Furthermore, a decentralized swarm of peers cannot provide reliable storage guarantees without a reliable replication mechanism allowing some nodes to go offline while guaranteeing that the data is always available.

Overall, the IPFS design creates a solid foundation for a sustainable, efficient and decentralized storage network. We hope that by highlighting the current challenges, we can help to address them in future releases of the protocol.

\begin{acks}
	The authors would like to thank the anonymous referees for their valuable comments and helpful suggestions.
    The authors express their gratitude to the contributors of passive DNS data to the European Data Sharing Collective --- SIE Europe.
	This work was supported by \grantsponsor{PL}{Protocol Labs}{https://research.protocol.ai/}
	under Grant No.~\grantnum{PL}{PL-RGP1-2021-054}. %
	This work was also supported by the \grantsponsor{DFG}{German Research Foundation (DFG)}{https://www.dfg.de/}
	within the Collaborative Research Center (CRC) \grantnum[https://gepris.dfg.de/gepris/projekt/210487104]{DFG}{SFB 1053: MAKI}.
    This work has been partially supported by the Cisco grant number 2020-216508 {Hybrid-ICN} Interoperability with {IPFS}. 
\end{acks}

\bibliographystyle{ACM-Reference-Format}
\bibliography{references}

\appendix
\section{Ethics}
\balance
The topology and traffic datasets raise limited ethical concerns. They involve collecting IP addresses, yet we do not attempt to map these back to personal identities, as such analysis was not within the scope of this study. The \bs and HydraBooster datasets additionally contain personal information, as they cover CID requests and advertisements from the clients. However, we do not trigger extra data collection. We anonymize IP addresses and do not perform lookups on the CIDs to infer the nature of the content exchanged.

We provide a user-facing website with our data management policy~\cite{trudi_privacy_policy}. We use collected data only for purposes that are permissible with respect to Art. 6 (4) GDPR. We operate within the bounds of European data protection law, specifically Art. 5 (1) (e) GDPR. We are storing the data for up to 24 months or until we do not need it anymore to complete our research, whichever comes sooner. We are constantly reflecting on our data storage practices. If we conclude that some data fields or whole data sets are no longer needed for our intended purpose, we will delete them accordingly.

Part of our research is based on active DNS scans and follows industry best practices in network measurements~\cite{menlo,considerations,zmap}. To distribute the load across different authoritative nameservers, we have implemented randomization in our input list of domain names, avoiding simultaneous scanning of all domains under the same TLD. We expect that a portion of our DNS requests has been resolved from the internal cache of Cloudflare Public DNS resolver.

\begin{sloppypar}
Furthermore, we tried to minimize the load caused by our research on the network. In IPFS, the provider records are stored on the 20 closest nodes to the $\text{CID}$ hash. The original $\text{FINDPROVIDERS(CID)}$ query performs a single DHT walk towards the $\text{CID}$ hash and asks all the encountered nodes on the path about all the providers for the target $\text{CID}$. By default, this walk ends when either enough provider records have been found or when all 20 closest peers to the $\text{CID}$ were queried. 
Our version of the $\text{FINDPROVIDERS(CID)}$ call terminates once all 20 closest peers to the target $\text{CID}$ are queried. In our analysis, we found that for 99.55\% of the CIDs, our modified $\text{FINDPROVIDERS(CID)}$ call discovered fewer than 20 providers and thus behaved exactly as the original $\text{FINDPROVIDERS(CID)}$ call. Consequently, for the vast majority of CIDs, our modification incurred no additional network overhead. The additional overhead incurred by the remaining 0.45\% of CIDs was limited to obtaining providers from at most 19 additional peers.
Therefore, our modifications to the $\text{FINDPROVIDERS}$ function were implemented to minimize any potential overhead and ensured that the vast majority of CIDs experienced no additional network burden.
\end{sloppypar}

For the IP geolocation, we used the MaxMind GeoIP database~\cite{maxmind}. We downloaded and later locally queried their GeoLite IP database, which operates entirely offline on our machines. No IP information left our local machine.

\end{document}